# Capturing nucleation at 4D atomic resolution

Jihan Zhou[1*], Yongsoo Yang[1*], Yao Yang[1], Dennis S. Kim[1], Andrew Yuan[1], Xuezeng Tian[1], Colin Ophus[2], Fan Sun[3], Andreas K. Schmid[2], Michael Nathanson[4], Hendrik Heinz[4], Qi An[5], Hao Zeng[3], Peter Ercius[2] & Jianwei Miao[1]

*[1]Department of Physics & Astronomy and California NanoSystems Institute, University of California, Los Angeles, CA 90095, USA. [2]National Center for Electron Microscopy, Molecular Foundry, Lawrence Berkeley National Laboratory, Berkeley, CA 94720, USA. [3]Department of Physics, University at Buffalo, the State University of New York, Buffalo, NY 14260, USA. [4]Department of Chemical and Biological Engineering, University of Colorado at Boulder, Boulder, CO 80309, USA. [5]Chemical and Materials Engineering, University of Nevada, Reno, NV 89557-0388, USA.*

*[*]These authors contributed equally to this work.*

**Nucleation plays a critical role in many physical and biological phenomena ranging from crystallization, melting and evaporation to the formation of clouds and the initiation of neurodegenerative diseases. However, nucleation is a challenging process to study in experiments especially in the early stage when several atoms/molecules start to form a new phase from its parent phase. Here, we advance atomic electron tomography to study early stage nucleation at 4D atomic resolution. Using FePt nanoparticles as a model system, we reveal that early stage nuclei are irregularly shaped, each has a core of one to few atoms with the maximum order parameter, and the order parameter gradient points from the core to the boundary of the nucleus. We capture the structure and dynamics of the same nuclei undergoing growth, fluctuation, dissolution, merging and/or division, which are regulated by the order parameter distribution and its gradient. These experimental observations differ from classical nucleation theory (CNT) and to explain them we propose the order parameter gradient (OPG) model. We show**

**the OPG model generalizes CNT and energetically favours diffuse interfaces for small**

**nuclei and sharp interfaces for large nuclei. We further corroborate this model using**

**molecular dynamics simulations of heterogeneous and homogeneous nucleation in**

**liquid-solid phase transitions of Pt. We anticipate that the OPG model is applicable to**

**different nucleation processes and our experimental method opens the door to study the**

**structure and dynamics of materials with 4D atomic resolution.**

Nucleation is a ubiquitous phenomenon in many scientific disciplines[1-3]. To study the nucleation mechanism, an ideal method would be to determine the 3D atomic or molecular structure of newly formed nuclei and monitor their dynamics. Although crystallography has long been used to determine the 3D atomic structure of molecules[4], it cannot be applied to study nucleation due to its requirement of a global average of many identical unit cells, whereas nuclei form locally and irregularly. Over the years, a number of experimental and computational methods have been implemented to investigate nucleation processes, such as x-ray scattering[5,6], electron microscopy[7-10], atomic force microscopy[11,12], atom probe tomography[13], video and confocal microscopy[14-16], molecular dynamics (MD) and Monte Carlo simulations[17-21], and others[1,22]. On the theoretical side, CNT has been the most widely used model to describe nucleation processes[1-3,23]. While CNT can explain many nucleation phenomena, in some cases its predicted nucleation rates can differ from the measured values by several orders of magnitude[1,24,25]. To alleviate these inconsistencies, non-classical nucleation theories have been proposed, including density functional theory[26,27,1], diffuse interface theory[28,29,1] and dynamical nucleation theory[30]. In recent years, a two-step nucleation model, stemming from computational and experimental observations, has been developed to describe crystallization in solution[17,24,25,31-33], where a cluster of solute molecules forms first, followed by the formation of a crystal nucleus inside the cluster. However, despite all these developments, it remained unachievable to experimentally determine the 3D atomic structure and dynamics of early stage nuclei to correlate with the nucleation theory.

Here, we implement atomic electron tomography (AET)[34], a method capable of determining the 3D atomic coordinates and composition of materials without the assumption of crystallinity[35,36], to study early stage nucleation dynamics in solid-solid phase transitions. By tracking the same FePt nanoparticles at multiple annealing times, we probe the dynamics of the 4D atomic structure of early stage nuclei. We observe that nucleation starts on the surface of the nanoparticles (i.e. heterogeneous nucleation) and capture the same nuclei undergoing growth, fluctuation, dissolution, merging and/or division. We discover that nucleation dynamics are regulated by the order parameter and its gradient in the nuclei. To explain these experimental observations, we propose the order parameter gradient (OPG) nucleation model of which CNT represents a special case. We show that the OPG model has lower nucleation energy barriers than that of CNT. We further corroborate the OPG model by performing MD simulations of early stage nucleation in liquid-solid phase transitions of Pt.

**Capturing 4D atomic motion with AET**

AET has been used to reveal the 3D atomic structure of dislocations, stacking faults, grain boundaries, chemical order/disorder and point defects, and determine the atomic displacement and strain tensor with high precision[34-40]. But all of these studies were of static structures. To probe the 4D atomic structure of early stage nucleation, we have tracked the same nuclei at different annealing times and applied AET to determine their 3D atomic positions and species at each time (Methods). We used FePt nanoparticles as a model system because binary alloys have been widely used to study phase transitions[2] and FePt is a very promising material for next generation magnetic recording media[36,41]. As-synthesized FePt nanoparticles form a chemically disordered face-centred cubic (fcc) structure (A1 phase)[41]. With annealing, the A1 phase undergoes a solid-solid transition to an ordered face-centred tetragonal ($L1_0$) phase or a chemically ordered fcc ($L1_2$) phase, depending on the chemical composition[36,41]. To validate the experimental method of 4D AET, we first performed a consistency check experiment of FePt nanoparticles undergoing phase transitions. We annealed the nanoparticles at 520°C for 9 minutes in vacuum and acquired two independent, sequential tilt series of an FePt nanoparticle (termed particle 1) using an annular dark-field scanning transmission electron microscope[42] (Methods and Extended Data Table 1). After reconstructing the two data sets using a GENeralized Fourier Iterative REconstruction algorithm (GENFIRE)[36,43], we located and identified all the individual Fe and Pt atoms without the assumption of crystallinity (Methods). Extended Data Figs. 1a-f show the 3D atomic models obtained from the two independent measurements of the same nanoparticle. By comparing their 3D atomic coordinates and chemical species, we confirmed that 95.4% of atoms are consistent between the two models and the precision of our 3D atomic structure determination method is 26 pm (Extended Data Fig. 1g).

Next, we trapped the same FePt nanoparticles at different annealing times and acquired a tilt series at each time (Methods). By applying the same reconstruction, atom tracing, atom identification and refinement procedures, we obtained a 3D atomic model for each tilt series. Figures 1a-c show the atomic models of the same nanoparticle (named particle 2) with an accumulated annealing time of 9, 16 and 26 minutes, respectively. We observed that the total number of atoms in the nanoparticle was slightly changed at the three annealing times (Extended Data Table 1). This was caused by atomic diffusion between nanoparticles during annealing, as confirmed by an energy-dispersive x-ray spectroscopy experiment (Extended Data Fig. 2). The overall 3D shape of the nanoparticle was similar between 9 and 16 minutes of annealing, but changed from 16 to 26 minutes. A fraction of the surface and sub-surface atoms were re-arranged to form $L1_0$ phases, but the Pt-rich core of the nanoparticle remained the same (Figs. 1d-f), which is evident by comparing the same internal atomic layers along the [010] direction (Figs. 1g-i). These experimental observations can be explained by vacancy-mediated atomic diffusion during annealing as it is energetically more favourable to create vacancies on or near the surface than in the core of the nanoparticle[2]. Extended Data Fig. 3 shows the 3D atomic models of another FePt nanoparticle (named particle 3) with an accumulated annealing time of 9 and 16 minutes, exhibiting similar results to Fig. 1.

**Revealing heterogeneous nucleation sites**

The annealed FePt nanoparticles consist of A1, L1$_0$, and L1$_2$ phases. We quantified these phases using the short-range order parameter (Methods), which for simplicity we term the order parameter throughout this article. Based on the order parameter, we identified nuclei with the L1$_0$, Fe-rich A1, Pt-rich A1, Fe-rich L1$_2$ and Pt-rich L1$_2$ phases in these nanoparticles (Methods), where we define a nucleus with a minimum of 13 atoms (the centre atom plus its 12 nearest fcc neighbours). As the L1$_0$ phase nuclei are more abundant than the L1$_2$ phase ones in the nanoparticles and the former is technologically more important[41], we focused on the analysis of the L1$_0$ phase nuclei in this work. Careful examination of all the nuclei indicates that each early stage nucleus has a core of one to few atoms with the maximum order parameter. To locate the nucleation sites, we searched for the cores of all the L1$_0$ phase nuclei inside the nanoparticles. The distribution of the nucleation sites in particle 1 is in agreement between two independent measurements (Extended Data Fig. 4a-c). Figure 2a-d and Extended Data Fig. 4d-f show the evolution of the nucleation sites as a function of the annealing time in particles 2 and 3, respectively. If the core of a nucleus is within one unit cell distance (3.87 Å) from the surface, we define it as a surface site. Otherwise, it is defined as a sub-surface site. Most nucleation sites in particles 2 and 3 are located on the facets, edges or corners, where the <100> and <111> facets are shown in magenta and green colour, respectively. Compared to particles 2 and 3, particle 1 has more nucleation sites at the sub-surface, because many of its nuclei are relatively large and extends further into the nanoparticle. All our observations confirm that the nucleation is heterogeneous, which is energetically more favourable than homogeneous nucleation[1-3].

**Capturing nucleation dynamics at 4D atomic resolution**

To probe early stage nucleation dynamics, we tracked the same nuclei in each particle at different annealing times (termed common nuclei). By quantitatively comparing all the nuclei of the same particle at different annealing times, we identified 33 and 25 common nuclei in particles 2 and 3, respectively (Methods). As each atom is associated with an order parameter, we define the effective number of atoms by summing up the order parameters in each nucleus. We found that the order parameter of the nucleus core ($\alpha_0$) is correlated with the effective number of atoms (Extended Data Fig. 4g). Based on the effective number of atoms, we divided the common nuclei into three groups: growing, fluctuating and dissolving nuclei (Methods). Figure 3 shows five growing, fluctuating and dissolving nuclei in particle 2, where each nucleus is represented by an atomic model and a 3D contour map with an order parameter equal to 0.7 (red), 0.5 (purple) and 0.3 (light blue). Particle 2 has 14 growing, 14 fluctuating and 5 dissolving nuclei (Fig. 3, Extended Data Figs. 5, 6, and 7) and particle 3 has 16 growing and 9 dissolving nuclei. Among these common nuclei, we also observed merging and dividing nuclei, shown in Fig. 3g-i, Extended Data Figs. 5b-d and 6e.

In addition to the effective number of atoms, we found that the OPG also plays an important role in nucleation dynamics, which points from the core of each nucleus to its boundary. Figure 4a-c show the order parameter distribution of a growing nucleus in particle 2 (Fig. 3a-c) along the [110], [111] directions and with radial average, respectively, where the order parameter increases with the increment of the annealing time. Figure 4d-f shows the 3D OPG distribution of the same nucleus at three different annealing times. As the nucleus grows, the OPG spreads out further along the radial direction. These observations are corroborated by the analysis of other growing, fluctuating and dissolving nuclei (Extended Data Fig. 8). To perform a quantitative analysis, we summed up the OPG inside each nucleus, which we term the effective surface area of the nucleus as it has the same dimension as area. Figure 2e shows a plot of the effective surface area vs. the effective number of atoms for all the nuclei in particles 2 and 3. The dissolving nuclei are clustered near the lower left corner of the plot, while both small and large nuclei can fluctuate as a function of time.

**The order parameter gradient model**

Our experimental study of early stage nucleation reveals three observations that cannot be explained by CNT[1,2,23]. First, early stage nuclei are anisotropic, as characterized with sphericity[44], a measure of how closely the shape of a 3D object approaches a perfect sphere. Extended Data Fig. 4h shows the sphericity of the nuclei as a function of the effective number of atoms, where the majority of the nuclei have a sphericity between 0.5 and 0.9 (with 1.0 as a perfect sphere). The nonspherical shape of early stage nuclei is caused by geometrical constraint, local inhomogeneity and anisotropy of the interfacial tension. This result is also consistent with the previous experimental observation of the nucleus shape of anisotropic molecules using atomic force microscopy[11]. Second, each nucleus has a core of one to few atoms with the maximum order parameter and the OPG points from the core to the boundary of the nucleus (Fig. 4d-f and Extended Data Fig. 4g), resulting in a diffuse interface between the nucleus and its parent phase. Third, we observed the same nuclei undergoing growth, fluctuation, dissolution, merging and/or division (Fig. 3 and Extended Data Figs. 5-7), which are regulated by the order parameter distribution and its gradient. To account for these experimental results, we propose the OPG nucleation model

$$\Delta G = -\Delta g \int \alpha(\vec{r}) \, dV + \int \gamma \left| \vec{\nabla} \alpha(\vec{r}) \right| dV \quad , \qquad (1)$$

where $\Delta G$ is the total free energy change, $\Delta g$ the free energy change per unit volume, $\alpha$ the order parameter distribution between 0 and 1, and $\gamma$ the interfacial tension of a sharp interface that can be anisotropic. Equation (1) is for homogeneous nucleation and for heterogeneous nucleation it is multiplied by a shape factor[1,2]. The first term in equation (1) stands for the effective volume energy change of a nucleus. The second term represents the effective interfacial energy of the nucleus, which can be derived from two independent methods (Extended Data Fig. 9a and Methods).

The OPG model reduces to CNT when the order parameter is represented by a sharp interface

$$\alpha(r') = \begin{cases} 1 & r' \le r \\ 0 & r' > r \end{cases} \quad . \qquad (2)$$

Substituting equation (2) into equation (1) gives the total free energy change in CNT[1,2,23]

$$\Delta G = -\frac{4\pi r^3}{3} \Delta g + 4\pi r^2 \gamma \quad . \qquad (3)$$

With spherical symmetry, we derive a general formula for determining the nucleation energy barrier of OPG and apply it to three specific diffuse interfaces (Methods). Extended Data Fig. 9b and c show that the nucleation energy barriers of the diffuse interfaces are lower than that of the sharp interface. Mathematically, we prove that if $\alpha$ monotonically decreases with radial distance, the OPG model always has lower nucleation energy barriers than CNT (Methods).

To apply the OPG model to the experimental data, we fit the order parameter distribution of each nucleus using a generalized Gaussian distribution, which can represent the nucleus with a smoothly varying boundary,

$$\alpha(r) = \alpha_0 e^{\left(-\frac{r}{\lambda}\right)^\beta} \quad , \qquad (4)$$

where $\alpha_0$, $\lambda$ and $\beta$ are the fitting parameters. Figure 4a-c and Extended Data Fig. 8 show the fit of equation (4) to the measured order parameter of several representative nuclei, indicating equation (4) (solid curves) is in good agreement with the experimental data (dots). According to equation (1), with any change of the order parameter and its gradient in a nucleus, the critical radius and the nucleation energy barrier are altered accordingly, creating a metastable state of the nucleus. Our experimental results indicate that early stage nuclei have various distributions of the order parameter and the gradient (Fig. 4 and Extended Data Fig. 8), resulting in different metastable states. When the difference between two order parameter distributions is small, the gap of the corresponding energy barriers is narrow, which facilitates the fluctuation of the nucleus between the two metastable states. Numerous such fluctuating nuclei were observed in our experimental data (Fig. 3d-l and Extended Data Fig. 6).

**Corroborating the OPG model with MD simulations**

To examine the applicability of the OPG model to different nucleation processes, we performed MD simulations of heterogeneous and homogeneous nucleation in liquid-solid phase transitions. The simulations were carried out using the large-scale atomic/molecular massively parallel simulator (LAMMPS)[45]. To enable cross-validation of the results, for heterogeneous nucleation we applied both the embedded-atom method potential and the interface force field to simulate two Pt liquid nanodroplets above the melting temperature (Methods). We then lowered the temperature to investigate early stage nucleation in crystallization. After analysing all the nuclei with the average local bond order parameter[46] (Methods), we found most nuclei are located on or near the surface of the two nanoparticles and each nucleus has a core of the maximum order parameter. Using the same criterion as the experimental data, we identified the common nuclei at different times and observed nucleation dynamics including growth, fluctuation, merging, division and dissolution. Figure 5a-d and Extended Data Fig. 10a-d show four representative growing, fluctuating, merging, dividing and dissolving nuclei for the embedded-atom method and the interface force field, respectively. The order parameter distributions of these nuclei with radial distance are shown in Fig. 5e-h and Extended Data Fig. 10e-h, indicating that nucleation dynamics is regulated by the order parameter distribution and its gradient. For homogeneous nucleation, we used the embedded-atom method potential with periodic boundary conditions to simulate a bulk Pt system undergoing liquid-solid phase transitions (Methods). Extended Data Fig. 11 shows four representative growing, fluctuating, merging, dividing and dissolving nuclei and their order parameter distributions. All the MD simulation results of heterogeneous and homogeneous nucleation are consistent with our experimental observations and further validate the OPG model.

## Discussion

By trapping the same nuclei at different annealing times, we applied AET to capture the structure and dynamics of the nuclei at 4D atomic resolution. We found that early stage nuclei are nonspherical, each nucleus has a core of one to few atoms with the maximum order parameter, and the OPG points from the core to the boundary. We captured the same nuclei undergoing growth, fluctuation, dissolution, merging and/or division. To explain these experimental observations, we proposed the OPG nucleation model, which was further corroborated by MD simulations of heterogeneous and homogeneous nucleation in liquid-solid phase transitions of Pt. There are several important implications of this work. First, the OPG model generalizes CNT and only reduces to CNT when the order parameter function is represented by equation (2). Furthermore, OPG solves an inconsistency problem in CNT[1,22]. For a single molecule, the first term in equation (3) of CNT is 0, and the second term is larger than 0, resulting in $\Delta G > 0$. But the OPG model resolves this inconsistency as both terms in equation (1) are 0 for a single molecule. Second, the competition between the effective volume energy change and interfacial energy terms in equation (1) creates the nucleation energy barriers. For small nuclei, the effective interfacial energy term dominates, creating lower energy barriers of diffusive interfaces than that of a sharp interface. For larger nuclei, the effective volume energy change term dominates, making the total free energy of the sharp interface decrease faster than that of diffuse interfaces (Extended Data Fig. 9c). Thus, the OPG model energetically favours diffuse interfaces for small nuclei and sharp interfaces for larger nuclei.

Third, according to CNT[1,2,23], the nucleation rate is proportional to $e^{-\frac{\Delta G^*}{k_B T}}$, where $\Delta G^*$ is the nucleation energy barrier, $k_B$ the Boltzmann constant and $T$ the temperature. Based on our experimental results and the OPG model, the order parameter distribution and its gradient of a nucleus can change as a function of time and each order parameter distribution gives a different nucleation energy barrier (i.e. a metastable state). Our experimental and MD results show that early stage nuclei can fluctuate between these metastable states. This may explain some of the discrepancies between experimentally measured nucleation rates and those predicted by CNT[1,24,25]. Fourth, by searching the literature, we found that the OPG model can be used to explain some other experimental and computer simulation results of nucleation processes. For example, the OPG model is consistent with the experimental measurements of the crystallization of a colloidal system[47] and computer simulations of gas-liquid and crystal nucleation using the Lennard-Jones potential[48,49]. As nucleation is such a widespread phenomenon in many different fields, it is impossible to explore all possible applications of the OPG model in a single paper. Nevertheless, all our experimental, MD simulation, and theoretical evidence indicate that that this model could be generally applied to different nucleation processes. Fifth, our experimental results on the early stage nucleation of the L1$_0$ FePt phase could expand our understanding of the critical conditions and requirements to make superior magnetic recording media and catalysts based on binary alloys[41,50]. Finally, all the seven experimental atomic models with 3D coordinates reported here will be deposited in the Materials Data Bank, an open database to serve the physical science community, which is analogous to the Protein Data Bank for the biological and life science communities. These experimentally measured coordinates can be used as direct input for density functional theory calculations and MD simulations of material properties[36], which is anticipated to open a new window to study the structure-property relationships of materials with 4D atomic resolution.

**Acknowledgements** We thank M. Pham, S. Osher, J. Rudnick, W. A. Goddard III, B. Wang, Y. Wang, A. Foi, L. Azzari and P. Sautet for stimulating discussions and T. Duden for his assistance with experiments. This work was primarily supported by STROBE: A National Science Foundation Science & Technology Center under Grant No. DMR 1548924. We also acknowledge the support by the Office of Basic Energy Sciences of the US DOE (DE-SC0010378) and the NSF DMREF program (DMR-1437263). The ADF-STEM imaging with TEAM 0.5 was performed at the Molecular Foundry, which is supported by the Office of Science, Office of Basic Energy Sciences of the U.S. DOE under Contract No. DE-AC02—05CH11231.

**Author information** The authors declare no competing financial interests. Correspondence and requests for materials should be addressed to J.M. (miao@physics.ucla.edu).

**Data availability** The raw and processed experimental data sets as well as all the Matlab source codes for the image reconstruction and data analysis approaches described in Methods will be freely available at www.physics.ucla.edu/research/imaging/OPG immediately after this work is published. The 3D atomic coordinates of particles 1, 2 and 3 at different annealing times will be deposited in the Materials Data Bank (www.materialsdatabank.org).

**Figures and Figure legends**

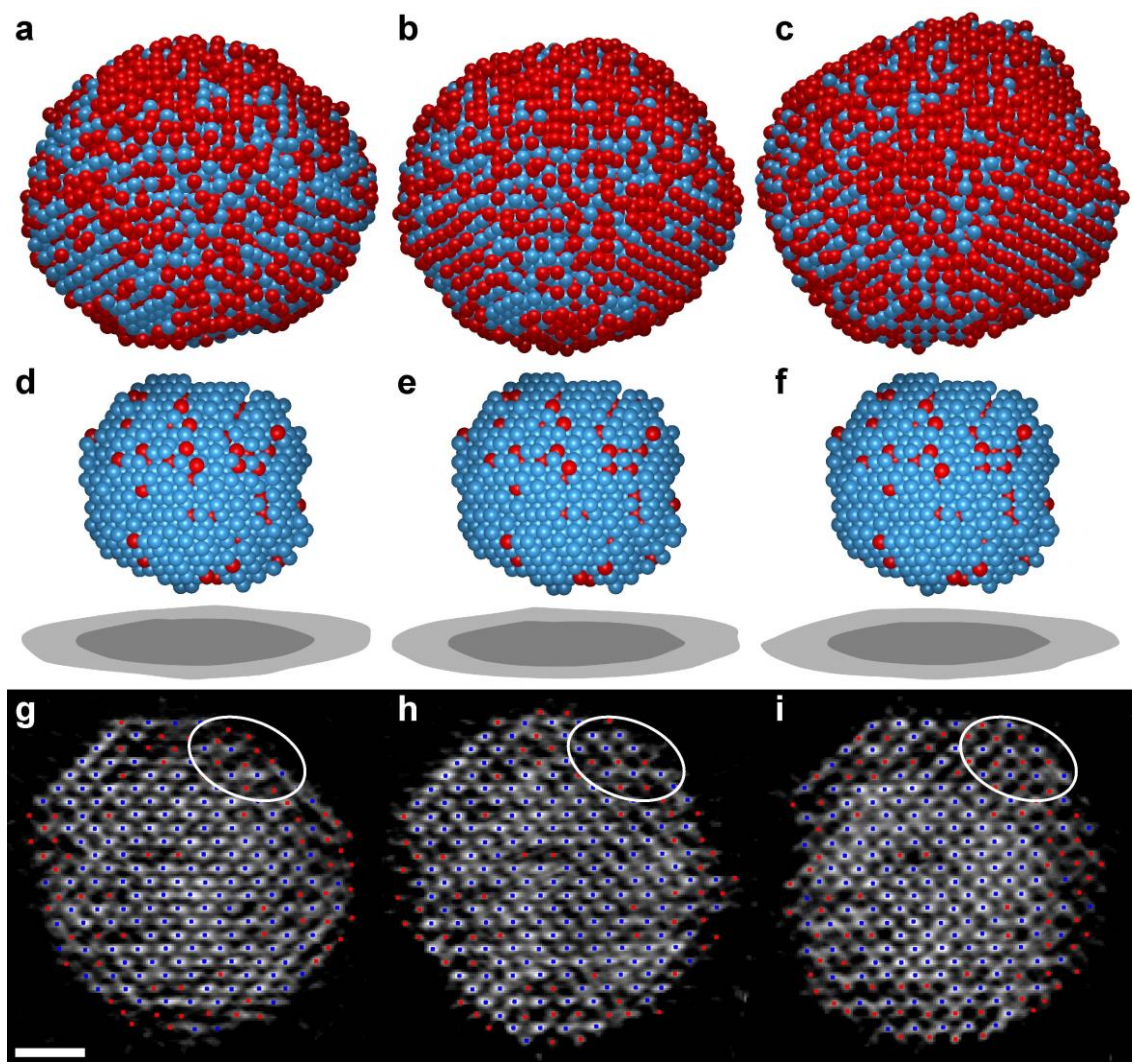

**Figure 1 | Capturing 4D atomic motion with AET**. **a-c,** 3D atomic models (Fe in red and Pt in blue) of an FePt nanoparticle with an accumulated annealing time of 9, 16 and 26 minutes, respectively. The 3D shape of the nanoparticle was similar from 9 to 16 minutes, but changed from 16 to 26 minutes. **d-f,** The Pt-rich core of the nanoparticle remained the same for the three annealing times. The light and dark grey projections show the whole nanoparticle and the core, respectively. **g-i,** The same internal atomic layer of the nanoparticle along the [010] direction at the three annealing times (Fe in red and Pt in blue), where a fraction of the surface and sub-surface atoms were re-arranged to form $L1_0$ phases (ellipses), but the Pt-rich core of the nanoparticle remained the same.

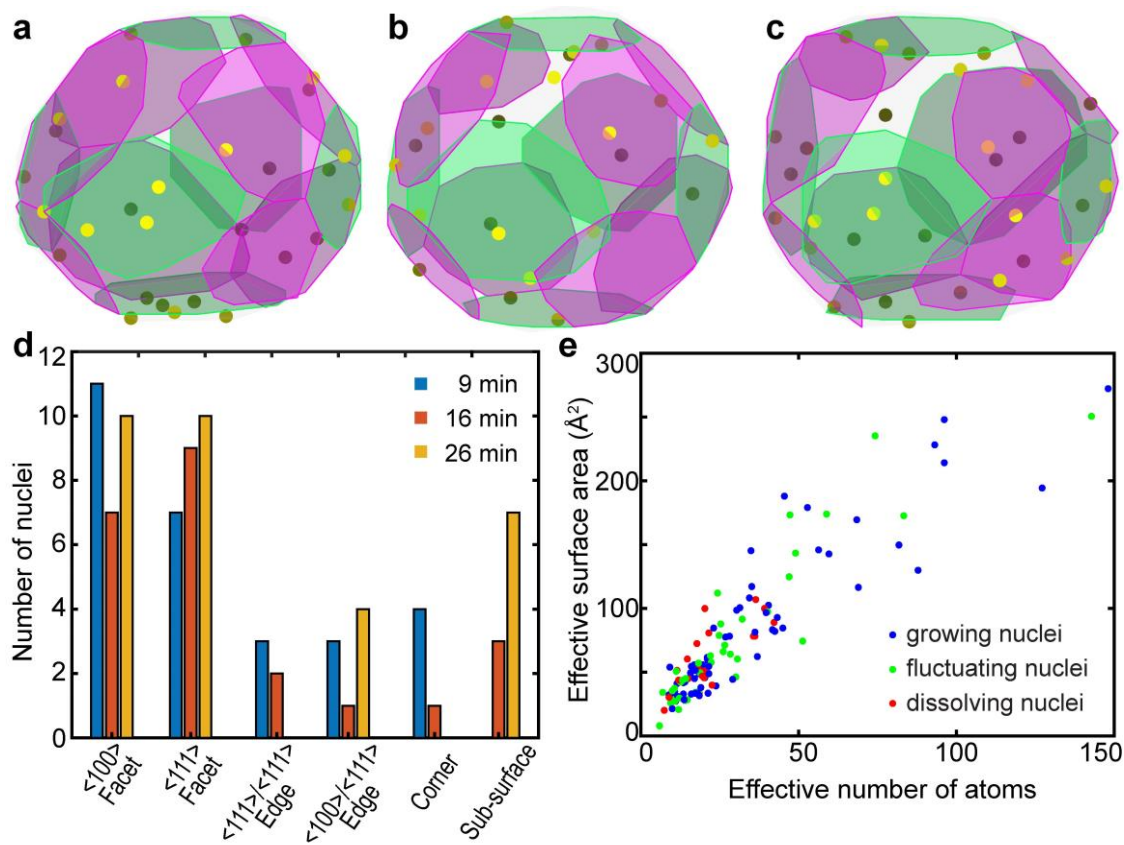

**Figure 2 | Revealing heterogeneous nucleation sites**. **a-c**, The distribution of the nucleation sites (circular dots) in particle 2 with an accumulated annealing time of 9, 16 and 26 minutes, respectively, where the lighter coloured dots are closer to the front side and the darker dots are closer to the back side of the nanoparticle. The <100> and <111> facets are in magenta and green, respectively. **d**, The histogram of the nucleation site distribution in particle 2, where most nucleation sites are located on the facets, edges or corners. **e**, A plot of the effective surface area vs. the effective number of atoms for all the nuclei in particles 2 and 3.

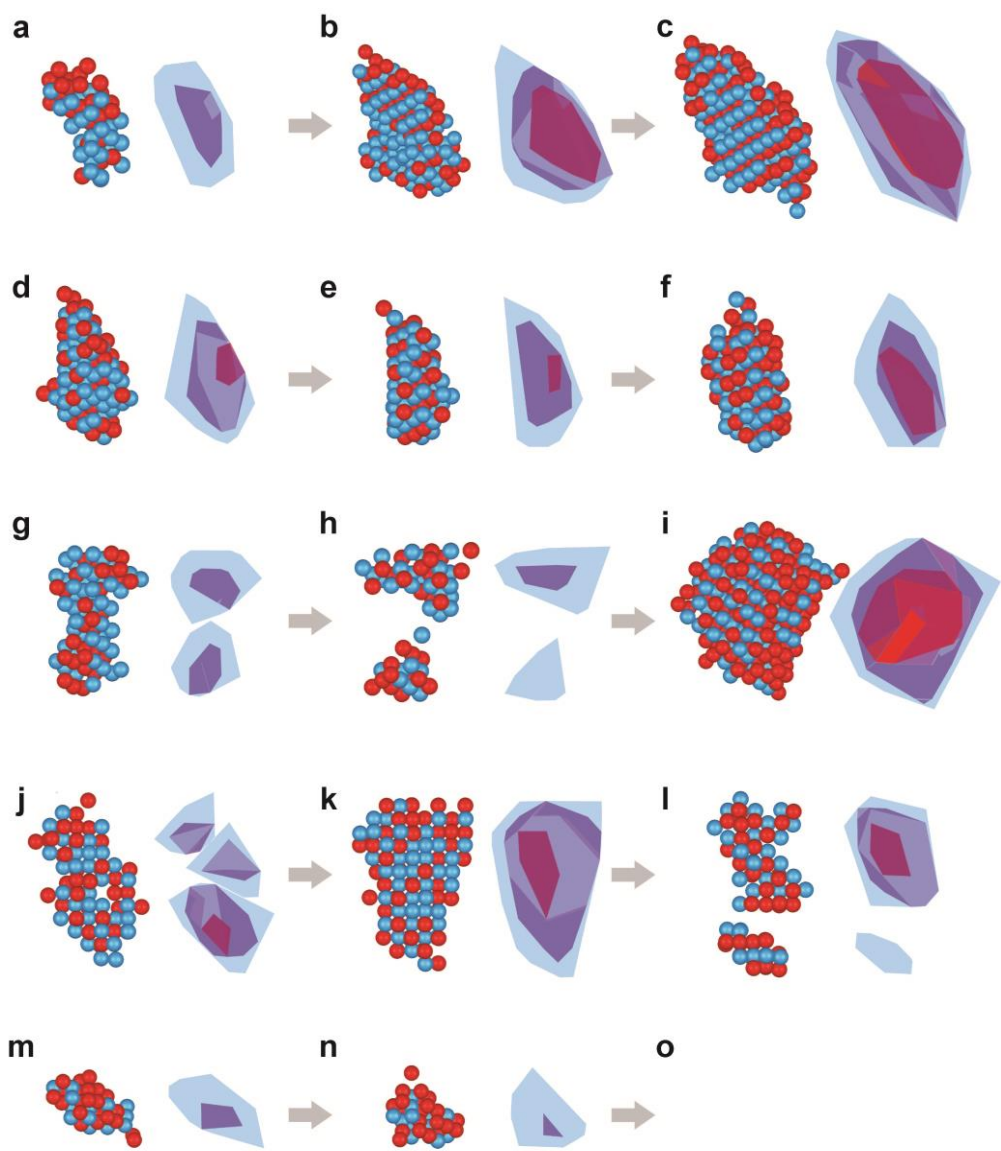

**Figure 3 | Experimental observation of the same nuclei undergoing growth, fluctuation, dissolution, merging and/or division at 4D atomic resolution. a-c**, A representative growing nucleus with an accumulated annealing time of 9, 16 and 26 minutes, respectively, where the atomic models show Fe (red) and Pt (blue) atoms with an order parameter ≥ 0.3 and the 3D contour maps show the distribution of an order parameter of 0.7 (red), 0.5 (purple) and 0.3 (light blue). **d-l**, Three representative fluctuating nuclei at three annealing times, including merging and dividing nuclei. **m-o**, A representative dissolving nucleus at three annealing times, which dissolved at 26 minutes (**o**).

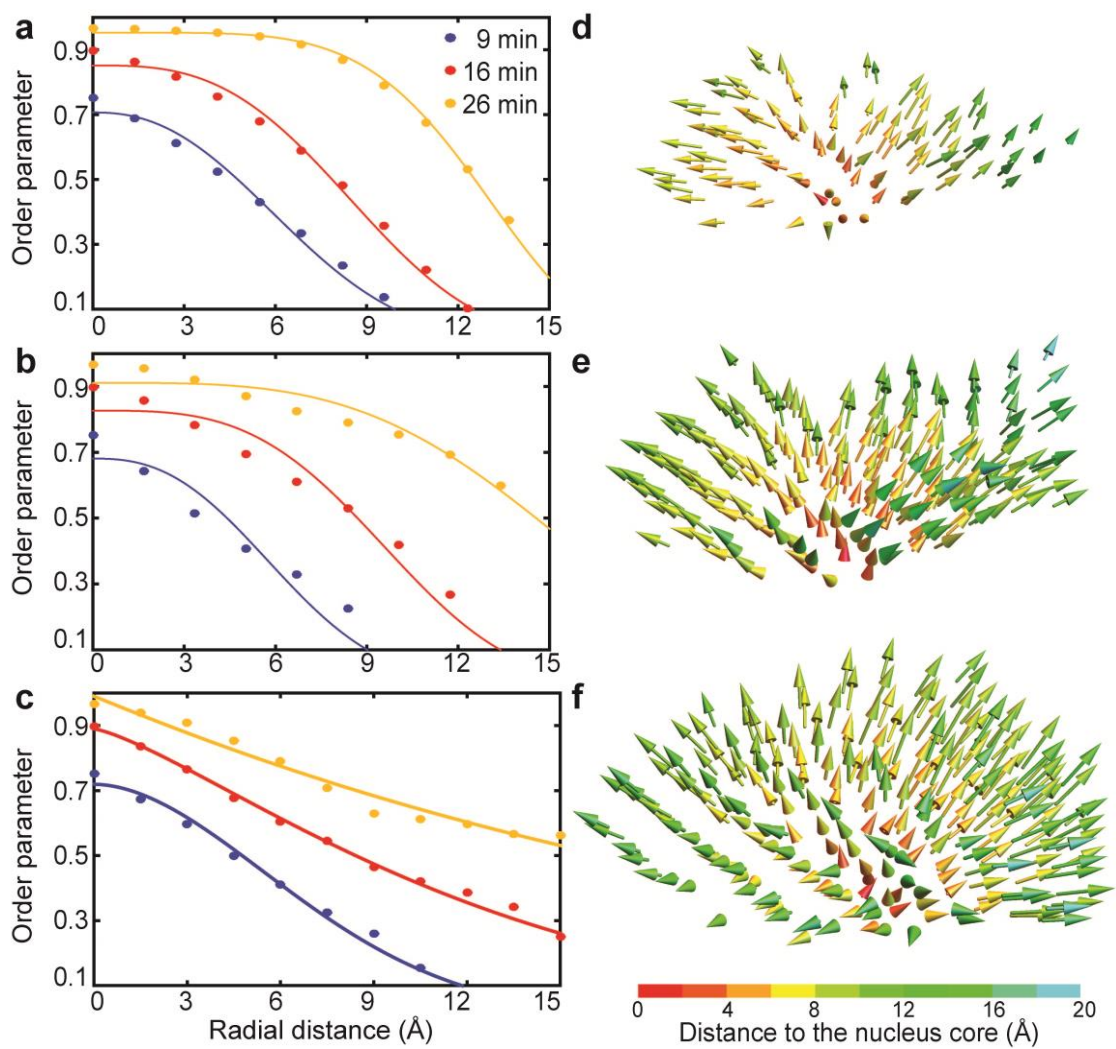

**Figure 4 | The 3D distribution of the order parameter and its gradient inside a representative nucleus**. **a-c**, The order parameter distribution of a growing nucleus (Fig. 3a-c) along the [110], [111] directions and with radial average, respectively, where the dots represent the experimental data and the curves are the fitted results with equation (4). **d-f**, The 3D OPG distribution of the nucleus at three annealing times, respectively, where the colours represent the distance to the nucleus core. With the growth of the nucleus, the OPG spreads out further along the radial direction.

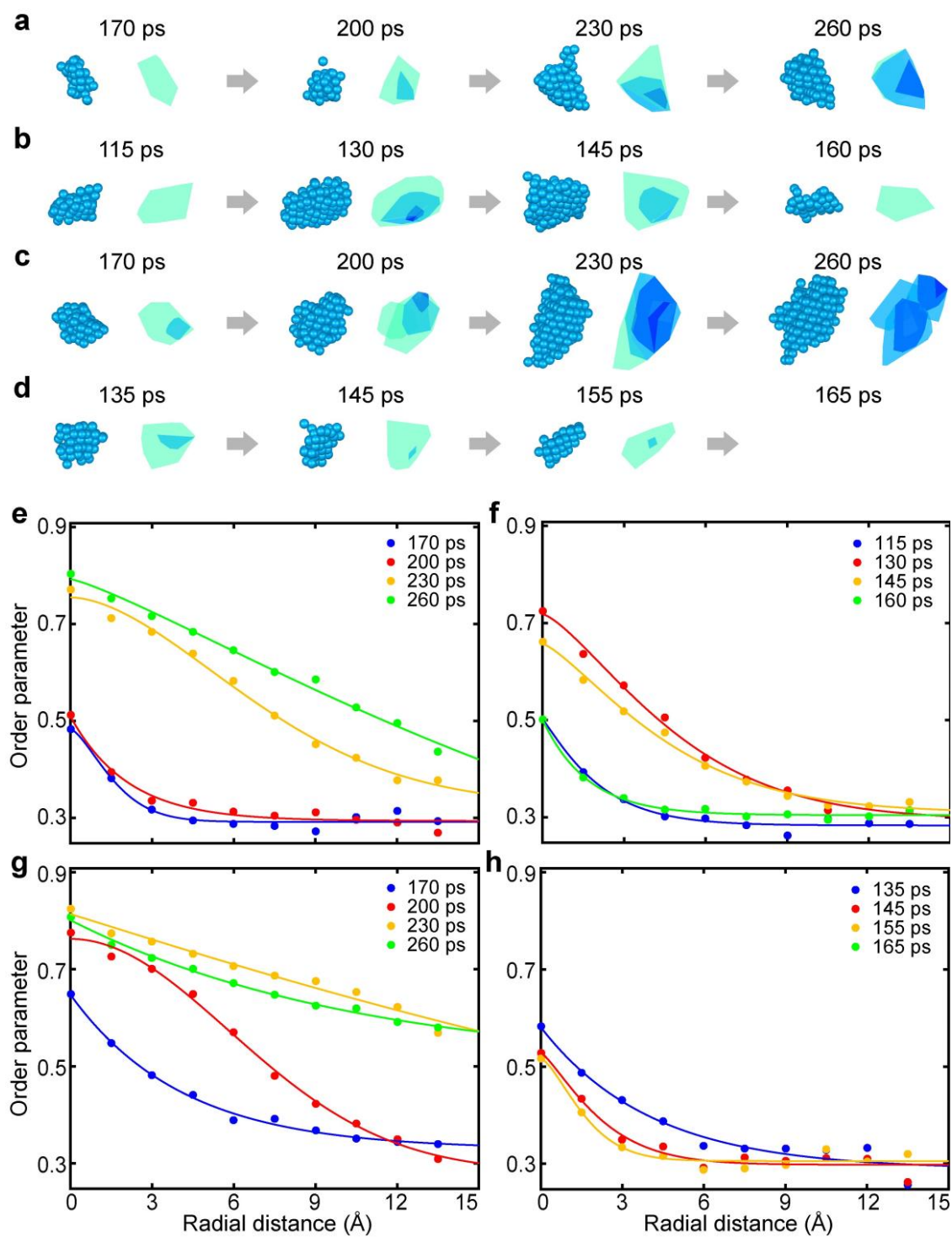

**Figure 5 | Nucleation dynamics in liquid-solid phase transitions of a Pt nanoparticle, obtained by MD simulations with the embedded-atom method potential**. **a**, A representative growing nucleus, where the atomic models show the Pt atoms with an order parameter ≥ 0.3 and the 3D contour maps show the distribution of an order parameter of 0.7

(dark blue), 0.5 (light blue) and 0.3 (cyan). **b** and **c**, Two representative fluctuating nuclei, where merging and dividing nuclei were observed in (**c**). **d**, A representative dissolving nucleus, which dissolved at 165 ps. **e-h**, Radial average order parameter distributions of the four nuclei shown in (**a-d**), respectively, where the dots were obtained by time-averaging ten consecutive MD snapshots with 1 ps time intervals and the curves are the fitted results using equation (4) with a constant background. The results indicate that nucleation dynamics are regulated by the distribution of the order parameter and its gradient.

## METHODS

**Data acquisition.** FePt nanoparticles were synthesized using the procedures published elsewhere[51]. After deposited on to 5-nm-thick silicon nitride membranes, the nanoparticles were annealed at 520 °C (below the melting temperature) for 9 minutes in vacuum. A set of tomographic tilt series were acquired from several FePt nanoparticles using the TEAM 0.5 microscope and the TEAM stage. Images were collected at 200 kV in ADF-STEM mode (Extended Data Table 1). To minimize sample drift, four to five images per angle were measured with 3 μs dwell time. For the consistency check experiment, we took a second set of tomographic tilt series from the same nanoparticles under the identical experimental conditions. For the dynamics study experiment, we took the nanoparticles out of microscope and annealed them at 520 °C for additional 7 minutes. Based on the pattern of the nanoparticle distribution on the substrate, we identified the same nanoparticles and acquired a second set of tomographic tilt series from them. We then annealed the same nanoparticles at 520 °C for additional 10 minutes and acquired a 3$^{rd}$ set of tilt series. We chose three FePt nanoparticles to present in this work. Particle 1 was annealed for 9 minutes and two independent, sequential tilt series were acquired under the same experimental conditions. Particle 2 was annealed with an accumulated time of 9, 16 and 26 minutes and a tilt series was taken at each time. Particle 3 was annealed with an accumulated time of 9 and 16 minutes and a tilt series was acquired at each time. To monitor any potential structural changes induced by the electron beam, we took 0° projection images before, during and after the acquisition of each tilt series and ensured that no noticeable structural changes were observed during the data acquisition for particles 1, 2 and 3. The total electron dose of each tilt series for particles 1, 2 and 3 was estimated to be between $7.6\times10^5$ e$^-$/Å$^2$ and $8.5\times10^5$ e$^-$/Å$^2$ (Extended Data Table 1), which is 5.6 to 6.3 times lower than that used in ref. 36.

**Image post-processing, denoising and GENFIRE reconstructions.** The four to five images acquired at each tilt angle were registered using normalized cross correlation[52] and then averaged. Linear stage drift at each tilt angle was estimated and corrected during the image registration. Scan distortion correction was also performed to correct for the imperfections in the calibration of the x- and y- scanning coils[35,36]. The experimental ADF-STEM images have mixed Poisson and Gaussian noise, and a sparse 3D transform-domain collaborative filtering[53] was applied to denoise the average image of each tilt angle. These post-processing and denoising methods have shown their robustness throughout other experimental data and multislice simulations[35,36].

After background subtraction and alignment, each tilt series was reconstructed using the GENFIRE algorithm[36,43]. From the initial 3D reconstruction, we applied the angular refinement routine implemented in GENFIRE to automatically correct the angular errors due to sample holder rotation and/or stage instability. After the automatic angular refinement, we manually applied additional angular correction and spatial alignment to minimize the distortions of Fourier space peak distributions and reduce the errors between the measured and calculated projections. After no further improvement can be made, we performed the final reconstruction of each tilt series using GENFIRE with the parameters shown in Extended Data Table 1.

**Determination of 3D atomic coordinates and species.** The 3D atomic coordinates and species of the nanoparticles were identified from the 3D reconstructions using the following procedure.

i) To enhance the tracing accuracy, we upsampled each 3D reconstruction by a factor of 3 using spline interpolation. All the local maxima were identified from the upsampled reconstruction.

ii) We implemented 3D polynomial fitting to localize the peak positions in each reconstruction, which generalizes a 2D method developed in particle tracking[54]. Starting from the highest-intensity local maximum peak, we cropped a $\sim 1.0 \times 1.0 \times 1.0$ Å$^3$ ($9 \times 9 \times 9$ voxel) volume with the selected local peak as the centre. We fit the volume with a 3D fourth-order polynomial function described elsewhere[54]. If a fitted peak position satisfied with a minimum distance constraint of 2 Å (i.e. the distance between two neighbouring atoms $\geq 2$ Å), we listed it as a potential atom position. According to our multislice simulations, the 3D polynomial fitting method is more accurate than 3D Gaussian fitting that has been used before[35,36].

iii) By applying the 3D polynomial fitting to all the identified local maxima, we obtained a list of potential atom positions. These positions were manually checked to correct for unidentified or misidentified atoms due to fitting failure or large chunk of connected intensity blobs from multiple atoms.

iv) We classified all the potential atoms into three different categories (non-atoms, potential Fe or Pt atoms) by applying an unbiased atom classification method described elsewhere[36]. With this classification procedure, we obtained an initial atomic model with 3D atomic coordinates and species from each 3D reconstruction.

v) Due to the missing wedge and experimental noise, there is local intensity variation in each 3D reconstruction. To further improve the atom classification accuracy, we performed local re-classification of the Fe and Pt atoms. For each atom in the initial atomic model, we drew a sphere with the atom as the centre and a radius of 6.76 Å. All the Fe and Pt atoms within the sphere were summed up to obtain an average Fe and Pt atom. The intensity distribution of the centre atom was compared with that of the average Fe and Pt atom. If the centre atom was closer to the average Fe than to the average Pt atom, it was assigned as an Fe atom, and vice versa. We iterated this process for all the atoms until the re-classification procedure was converged. Note that this process did not converge if the radius of the sphere was too small, and it became less effective if the radius was too large. By testing different radii, we found an optimal radius of 6.76 Å for this re-classification procedure.

**Refinement of 3D atomic coordinates and species.** We compared two atomic models of the same nanoparticle with each other. For particles 1 and 3, the two atomic models obtained from two experimental tilt series were compared. For particle 2, the 9-minute and 16-minute atomic models, and then the 16-minute and 26-minute atomic models were compared, respectively. We identified pairs of atoms (i.e. one atom from each model to form a pair), whose distance is within the radius of the Fe atom (1.4 Å). While the majority of the atom pairs have the same atomic species, there are a small percentage of atom pairs with different species. We developed the following atom flipping procedure to re-examine the atomic species of the small percentage of atom pairs.

i) An atom was randomly selected from the small percentage of atom pairs with different species. The projection intensities were calculated for all the tilt angles by flipping the selected atom (Fe to Pt or Pt to Fe), and the error between the calculated and measured projections was estimated. As flipping a single atom only affects a small local region of a projection, we only considered the local region in this process to increase the computational speed.

ii) If the error decreased after flipping, the flipped atomic species was updated in the model, otherwise the model was unchanged.

iii) Steps i) and ii) were repeated for all the small percentage of atom pairs and an updated atomic model was obtained. A global scale factor was calculated to minimize the error between the measured and calculated projections.

iv) Steps i)-iii) were iterated for all the small percentage of atom pairs until there was no change in the atomic species. This atom flipping method successfully converged for all datasets that we studied in this work.

From the updated atomic models, integrated intensity histograms for all atoms were plotted for each of the two atomic models in comparison. A double Gaussian function was fitted to the intensity histogram to identify obvious Fe atoms (integrated intensity smaller than the Fe atom peak), obvious Pt atoms (integrated intensity larger than Pt atom peak), and borderline atoms near the overlapping region of two Gaussians. We manually examined every borderline atom and its paired atom in the comparison model. If the paired common atom is classified as an obvious Fe or Pt atom, the atomic species of the borderline atom was re-classified to be consistent with its paired common atom.

After updating the chemical species for the atomic models in comparison, we refined the 3D atomic coordinates to minimize the error between the calculated and measured projections using the procedure described elsewhere[36]. During the refinement, we monitored both the total embedded-atom potentials and the root-mean-square deviation (RMSD) of the atomic coordinates between the atom pairs of the two models. For the RMSD calculation, appropriate affine transformations were applied to the atomic models to correct for remnant distortions. The iterative refinement process was terminated when a minimum RMSD was reached.

After finalizing the 3D coordinates, all the atomic species of unpaired atoms or paired atoms with different species in each model were refined again using steps i)-iv) described above. These atoms could be classified as Fe, Pt or non-atoms. To minimize misidentification, the atoms previously identified as obvious Pt atoms remained unaltered, and the atoms previously identified as obvious Fe atoms were prohibited from being identified as Pt atoms. Using this refinement procedure, we obtained the final refinement results of the seven atomic models with 3D atomic coordinates and species (Extended Data Table 1).

**Order parameter determination and nuclei identification.** The short-range order parameters of the atomic sites in the final atomic models were calculated for all 16 possible ordered phases from the FePt fcc lattice[55,56] (four $FePt_3$ $L1_2$, four $Fe_3Pt$ $L1_2$, six FePt $L1_0$, a Pt-rich A1, and a Fe-rich A1 phase). An order parameter $S_j$ for a given phase $j$ measures how many atomic sites in a set match the phase in question, normalized to the mean composition $f_{all}$ of all atomic sites. We define $f_j$ as the fraction of atomic sites in a given region that match phase $j$. The general expression relating the number of atomic sites with the correct composition $f_{rand}$ between phase $j$ and a disordered matrix by chance for a binary alloy is

$$f_{\text{rand}} = 2f_j f_{\text{all}} - f_j - f_{\text{all}} + 1 \quad . \qquad (5)$$

For a local measurement of the number of atomic sites matching with phase $j$ defined as $f$, the normalized order parameter $S_j$ is given by

$$S_j = \frac{f - f_{\text{rand}}}{1 - f_{\text{rand}}} = \frac{f - 2f_j f_{\text{all}} + f_j + f_{\text{all}} - 1}{f_j + f_{\text{all}} - 2f_j f_{\text{all}}} . \qquad (6)$$

This normalization step sets $S_j = 1$ for a perfectly chemically ordered set of atoms, and $S_j = 0$ if the fraction of atomic sites match phase $j$ by chance in a fully chemically disordered structure. Note that $S_j < 0$ can occur for phases with anti-correlated site compositions. We then applied a Gaussian kernel with a standard deviation of 0.75 fcc unit cells to prevent false positive grains at the disordered grain boundary, which occupy a small number of atomic sites. After determination of order parameters for all phases, every atom was assigned to one of the 16 phases based on its highest order parameter.

The nuclei in each atomic model were identified with the following procedure. For every atomic site, a sphere was drawn with the selected atom as the centre and a radius of 3.87 Å (one FePt fcc unit cell length). All the atomic sites inside the sphere were identified, which have the same ordered phase as that of the centre atom. If the highest order parameter atom inside the sphere is the centre atom, then the atom was defined as a core atom of a nucleus. Otherwise, the centre atom was tagged to be in the same nucleus as the highest order parameter atom, and a new sphere with the same radius and the highest order parameter atom as the centre was drawn to repeat the procedure until a nucleus core site was found. Applying this procedure to all atoms in each atomic model resulted in clusters of atoms with each cluster having a core. A cluster with a minimum of 13 atoms and order parameter $\geq 0.3$ was defined as a nucleus in this study. We chose a minimum of 13 atoms because an fcc cluster consists of a centre atom and 12 nearest-neighbour atoms. After identifying all the nuclei in the nanoparticles, we counted the number of atoms in the core of each nucleus. Using the criterion that the atoms in a nucleus core must be within top 95% of the maximum order parameter, we estimated that the core of each early stage nucleus has one to few atoms.

**Identification of common nuclei.** The nucleation dynamics study was performed on particles 2 and 3, which have three and two annealing times, respectively. To identify the common nuclei for particles 2 and 3 at different annealing times, we used the following three criteria. First, a common nucleus can form, grow, merge, divide or dissolve at any annealing time. Second, if a common nucleus exists in at least two different annealing times, each must overlap with at least another nucleus with more than 50% of the volume of the smaller nucleus. Third, a common nucleus must not overlap with any non-common nuclei at different annealing times with more than 50% of the volume of the smaller one. Based on these three criteria, we found 33 common nuclei for particle 2, including 14 growing, 5 dissolving and 14 fluctuating nuclei. For particle 3, we found 25 common nuclei with 16 growing and 9 dissolving one. Since particle 3 has only two annealing times, it cannot be used to identify fluctuating nuclei. For all the nuclei in the two particles, we also performed an analysis of the tetragonal distortion of the L1$_0$ phase. We obtained the c/a ratio and calculated a weighted mean and standard deviation with the number of atoms of each nucleus as a weight factor. The weighted c/a ratios are $0.98 \pm 0.02$, $0.98 \pm 0.03$ and $0.97 \pm 0.03$ for the three different annealing times of particle 1, and $0.97 \pm 0.02$ and $0.98 \pm 0.02$ for the two different annealing times of particle 2, respectively, which agree with the c/a ratio of 0.96 for the bulk L1$_0$ phase.

**Derivation of the OPG nucleation model**. In the OPG model, each atom or molecule in a nucleus is assigned with an order parameter ($\alpha$) between 0 and 1. By summing up the order parameter for all the atoms, the first term in equation (1) represent the effective volume energy difference of the nucleus. For example, for an atom with $\alpha = 0.6$, its contribution to the effective volume energy difference is *-0.6$\Delta g$ \*$\Delta V$*, where $\Delta V$ is the volume occupied by the atom. To derive the second term in equation (1), we divide a nucleus into many very small volumes (Extended Data Fig. 9a). The direction of the OPG inside each volume is along the $\Delta \vec{s}$ direction and the magnitude of the OPG is $\frac{|\Delta \alpha|}{\Delta d}$, where $\Delta \alpha$ is the order parameter difference in the volume and $\Delta d$ is the distance along the OPG direction. The interfacial tension of this volume is calculated by

$$\frac{|\Delta \alpha|/\Delta d}{1/\Delta d}\, \gamma = |\Delta \alpha|\, \gamma \qquad (7)$$

where $\gamma$ is the interfacial tension of a sharp interface with $\Delta \alpha = 1$. The total interfacial energy of the nucleus is obtained by adding up the interfacial energy of all the small volumes in the nucleus

$$\int |\Delta \alpha|\, \gamma\, ds = \int \gamma \left|\frac{\Delta \alpha}{\Delta d}\right|\, dV = \int \gamma |\vec{\nabla}\alpha|\, dV \ , \qquad (8)$$

which is the second term in equation (1).

Next, we provide a more mathematically rigorous proof of the effective interfacial energy of the OPG model. According to the Coarea formula of geometric measure theory[57], for a 3D scalar field $\Phi$ with the $C^1$ continuity conditions, integrating a function ($f$) over the isolevel $c$ in a region $\Omega$ is equivalent to doing a volume integral over $\Omega$,

$$\int_{-\infty}^{\infty}\left[\int_{\{\Phi(\vec{r}_c)=c\}\cap\Omega} f(\vec{r}_c)d\sigma\right]dc = \int_{\Omega} f(\vec{r})|\vec{\nabla}\Phi|dV \qquad (9)$$

where $\{\Phi(\vec{r}_c) = c\}$ is the level surface set containing all points $\vec{r}_c$ such that $\Phi(\vec{r}_c) = c$, $\int d\sigma$ is the surface integral, and $\{\Phi(\vec{r}_c) = c\}\cap\Omega$ is the intersection between $\{\Phi(\vec{r}_c) = c\}$ and $\Omega$. By choosing $f(\vec{r}) = \gamma$, equation (9) becomes

$$\gamma \int_{-\infty}^{\infty}\left[\int_{\{\Phi(\vec{r}_c)=c\}\cap\Omega} d\sigma\right]dc = \gamma \int_{\Omega} |\vec{\nabla}\Phi|dV \ . \qquad (10)$$

By substituting $\Phi(\vec{r}) = \alpha(\vec{r})$ with $0 \le \alpha \le 1$ into equation (10), we obtain

$$\gamma\, A_{eff} = \gamma \int_{0}^{1}\left[\int_{\{\alpha(\vec{r}_c)=c\}\cap\Omega} d\sigma\right]dc = \gamma \int |\vec{\nabla}\alpha(\vec{r})|dV \ . \qquad (11)$$

where $A_{eff}$ represents the effective surface area of $\alpha(\vec{r})$ and $\gamma\, A_{eff}$ is the effective interfacial energy of the OPG model, i.e. the second term in equation (1).

Finally, we want to make a distinction between the OPG model and the square-gradient model[1,58-60]. The square-gradient model consists of a $\left(\vec{\nabla}\rho\right)^2$ term, where the density ($\rho$) is represented as the order parameter. There are three differences between the square-gradient and the magnitude-gradient term, $|\vec{\nabla}\alpha|$, in the OPG model.

i) The square-gradient term is derived from the Taylor expansion as the gradient term goes to 0 after summing it along all the directions [1,58-60]. But the magnitude-gradient term can be rigorously derived from the Coarea formula of geometric measure theory[57], i.e. equations (9) – (11).

ii) Although the squared-gradient and the magnitude-gradient terms can both use the order parameter[1,61,62], the coefficients of the two terms are different. The coefficient of the square-gradient term requires a complicated integration[1,60]. In contrast, the coefficient of the magnitude-gradient term is $\gamma$, which has a clear physical meaning - the interfacial tension of a sharp interface between $\alpha = 1$ and 0.

iii) When substituting the order parameter distribution of the Heaviside step function (equation (2)) into the magnitude-gradient term in equation (1), the OPG model reduces to CNT. Therefore, OPG generalizes CNT and CNT represents a special case of OPG. Energetically, the OPG model favours diffuse interfaces for small nuclei and sharp interfaces for large nuclei (Extended Data Fig. 9c). In contrast, the square-gradient model cannot be reduced to CNT when substituting the Heaviside step function into the model.

**The energy barrier of the OPG model**. To determine the energy barrier of the OPG model in three dimensions, we re-write equation (1) with spherical symmetry

$$\Delta G = -4\pi \Delta g \int_0^r \alpha(r,r')\, r'^2 dr' + 4\pi\gamma \int_0^r \left|\vec{\nabla}'\alpha(r')\right| r'^2 dr' \qquad (12)$$

where $\alpha(r,r') = 0$ for $r' > r$. Using the Leibniz integral rule, we have

$$\frac{d(\Delta G)}{dr} = -4\pi\Delta g \frac{\partial}{\partial r}\left[\int_0^r \alpha(r,r')\, r'^2 dr'\right] + 4\pi\gamma \frac{\partial}{\partial r}\left[\int_0^r \left|\vec{\nabla}'\alpha(r,r')\right| r'^2 dr'\right]$$

$$= -4\pi\Delta g\left[\alpha(r,r)\, r^2 + \int_0^r \frac{\partial\alpha(r,r')}{\partial r}\, r'^2 dr'\right] + 4\pi\gamma\left[\left|\vec{\nabla}'\alpha(r,r)\right| r^2 + \int_0^r \frac{\partial\left|\vec{\nabla}'\alpha(r,r')\right|}{\partial r}\, r'^2 dr'\right]$$

$$= 4\pi r^2\left[\gamma\left|\vec{\nabla}\alpha(r,r)\right| - \Delta g\,\alpha(r,r)\right] + 4\pi\left[\gamma \int_0^r \frac{\partial\left|\vec{\nabla}'\alpha(r,r')\right|}{\partial r}\, r'^2 dr' - \Delta g \int_0^r \frac{\partial\alpha(r,r')}{\partial r}\, r'^2 dr'\right]$$

$$= 0 \qquad (13)$$

We re-arrange equation (13),

$$\Delta g\,\alpha(r,r) - \gamma\left|\vec{\nabla}\alpha(r,r)\right| = \frac{1}{r^2}\int_0^r \frac{\partial}{\partial r}\left[\gamma\left|\vec{\nabla}'\alpha(r,r')\right| - \Delta g\,\alpha(r,r')\right] r'^2 dr' \qquad (14)$$

where $\vec{\nabla}\alpha(r,r) \equiv \left.\frac{\partial\alpha(r,r')}{\partial r'}\right|_{r'=r}$. Equation (14) represents a general formula for determining the nucleation energy barrier of 3D systems, which can be solved numerically. Note that for 1D systems, because both the first and second terms in equation (1) are proportional to $r$, the OPG model has no energy barrier, which has been experimentally validated[63].

For a given $\alpha$, we can calculate the corresponding critical radius and energy barrier. Let's consider four specific cases of $\alpha$.

i) A sharp interface of the Heaviside step function (Extended Data Fig. 9b, black curve). Substituting equation (2) into (14), we have

$$r_c^* = \frac{2\gamma}{\Delta g} \quad , \qquad (15)$$

where $r_c^*$ is the critical radius. We substitute equation (15) into equation (12) to calculate the nucleation energy barrier of the sharp interface of CNT

$$\Delta G^* = \frac{16.7\alpha_0 r^3}{\Delta g^2} . \qquad (16)$$

ii) A diffuse interface with a linear decrease of $\alpha$ (Extended Data Fig. 9b, red curve). The order parameter distribution is represented by

$$\alpha = \alpha_0 \left( 1 - \frac{r'}{r} \right). \qquad (17)$$

Substituting equation (17) into equation (14) and then inserting the critical radius into equation (12), we obtain

$$r_c^* = \frac{2.7\gamma}{\Delta g} \qquad\qquad \Delta G^* = \frac{9.9\alpha_0\gamma^3}{\Delta g^2} . \qquad (18)$$

iii) A diffuse interface with a parabolic decrease of $\alpha$ (above the linear line in Extended Data Fig. 9b, blue curve). The order parameter distribution is specified by

$$\alpha = \alpha_0 - \alpha_0 \left( \frac{r'}{r} \right)^2 . \qquad (19)$$

From equations (19), (14) and (12), we calculate the critical radius and energy barrier

$$r_c^* = \frac{2.5\gamma}{\Delta g} \qquad\qquad \Delta G^* = \frac{13.1\alpha_0\gamma^3}{\Delta g^2} . \qquad (20)$$

iv) A diffuse interface with a parabolic decrease of $\alpha$ (below the linear line in Extended Data Fig. 9b, green curve). The order parameter is represented by

$$\alpha = \alpha_0 \left( \frac{r'}{r} - 1 \right)^2 . \qquad (21)$$

The corresponding critical radius and energy barrier are

$$r_c^* = \frac{3.3\gamma}{\Delta g} \qquad\qquad \Delta G^* = \frac{7.8\alpha_0\gamma^3}{\Delta g^2} . \qquad (22)$$

Extended Data Fig. 9c shows the total free energy changes as functions of the radial distance for the sharp and three diffuse interfaces. With a small radius, the interfacial energy term (the second term in equation (1)) dominates, creating a lower nucleation energy barrier of the diffusive interface than that of the sharp interface. With a large radius, the volume energy change term (the first term in equation (1)) dominates, making the total free energy of the sharp interface decrease faster than those of the diffuse interface.

Next, we prove that for a monotonically decreasing order parameter distribution, $\alpha$, the OPG model has lower energy barriers than CNT for both homogeneous and heterogeneous nucleation. Mathematically, it is equivalently to show that, if the volume energy difference term remains the same for the OPG model and CNT,

$$4\pi \int_0^r \alpha(r,r') \, r'^2 dr' \ = \ \frac{4\pi}{3} r^3 \ , \qquad (23)$$

then the interfacial energy term of OPG is always lower than or equal to that of CNT,

$$4\pi \int_0^r \left| \vec{\nabla}' \alpha(r,r') \right| r'^2 dr' \ \le \ 4\pi r^2 \ . \qquad (24)$$

As $\alpha$ monotonically decreases with radial distance, we have $\left| \vec{\nabla}'\alpha \right| = -\alpha'$. Substituting this relation and equation (23) into equation (24), we just need to prove

$$\left[ \int_0^r (-\alpha') \, r'^2 dr' \right]^{3/2} \le 3 \int_0^r \alpha(r,r') \, r'^2 dr' . \qquad (23)$$

Using Jensen's inequality[64] and integral by parts, we have

$$\left[\int_0^r (-\alpha')\, r'^2 dr'\right]^{3/2} \leq \int_0^r (-\alpha')\, (r'^2)^{3/2} dr' = 3\int_0^r \alpha(r,r')\, r'^2 dr' . \qquad (24)$$

The proof is for homogeneous nucleation. For heterogeneous nucleation we multiply the total free energy change by a shape factor and the remaining derivation is the same. Thus, we prove that for monotonically decreasing $\alpha$, the OPG model is energetically more favourable than CNT.

**MD simulations of heterogeneous and homogeneous nucleation in liquid-solid phase transitions of Pt.** To further validate the OPG model, we performed MD simulations on two Pt nanoparticles and a Pt bulk system using the LAMMPS package[46]. We first used an embedded-atom method potential to simulate a Pt nanoparticle of 32,000 atoms[65], which was put in a much larger box so that it does not interact with its periodic images. The nanoparticle was melt and equilibrated at 2,500 K and then quenched to room temperature with a cooling rate of 1 K·ps$^{-1}$. The heterogeneous crystal nucleation initiates at 1,050 K in the supercooling region. The potential energy significantly drops when crystallization initiates. To examine the detailed nucleation processes, we selected the cooling snapshot at 1,100 K and performed fixed temperature simulations at the 1,100 K using the NVT ensemble (constant number of particle, constant volume and constant temperature). Since the system was in supercooling region, the crystallization started at ~150 ps and ended at ~300 ps.

To cross-validate the MD results, we simulated another Pt nanoparticle of 13,500 atoms in the canonical (NVT) ensemble in LAMMPS using the interface force field as the interatomic potential[66]. The nanoparticle was melt at 2,750 K for 300 and the temperature was lowered to 2,000 K for 200 ps. At this temperature the Pt nanodroplet showed no nucleation. The nanodroplet was then quenched to 1,650 K for 1 ns with a cooling rate of 1.65 K·ps$^{-1}$. During this cooling period, nucleation and liquid-solid phase transitions of Pt were induced and observed. Coordinates were recorded every 1 ps during this part of the simulation and used to analyse the in-situ change of the order parameter and atomic displacements during the nucleation process.

In addition to heterogeneous nucleation, we also performed MD simulations of homogeneous nucleation using a bulk Pt system. An embedded-atom method potential was used to simulate 32,000 Pt atoms[65] and periodic boundary conditions were applied along three directions to eliminate the surface effects. The system was equilibrated at 2,500 K and quenched to room temperature with a cooling rate of 1 K·ps$^{-1}$. In contrast to the Pt nanodroplet, the bulk system crystallized at ~750 K during quench process, which is lower in temperature than the heterogeneous nucleation process. This is because the homogeneous system has much less nucleation sites than the nanodroplet. The nucleation process was examined at a fixed temperature of 800 K using the NPT ensemble (constant number of particle, constant pressure and constant temperature). The crystallization initiated in the first few picoseconds and ended at ~200 picoseconds.

**Order parameter definition and nuclei identification for the MD simulation results.** The order parameters of the Pt atoms in the MD simulations were calculated using local bond-orientation order parameter method[47,67,68]. The $Q_4$ and $Q_6$ order parameters were calculated up to the second shell with the shell radius of 3.8 Å as described elsewhere[68]. The order parameter was normalized between 0 and 1 where 0 corresponds to $Q_4=Q_6=0$ and 1 represents a perfect Pt fcc structure. To identify the nuclei formed during the heterogeneous and homogeneous nucleation, we applied the same method described above with a 4-Å-radius sphere and a minimum of 31 atoms. Common nuclei at different time points were also identified using the same method described above. Note that the local bond-orientation order parameter has been previously used to study crystallization with computer simulations[19,20].

To examine the 3D shapes of the nuclei in the MD simulations of heterogeneous and homogeneous nucleation, we calculated the sphericity of the nuclei in the crystallization of Pt (Extended Data Fig. 12). The distribution of the sphericity of the nuclei in the MD simulations is in good agreement with that of the experimental data (Extended Data Fig. 4h). In particular, for homogeneous nucleation we used the embedded-atom method potential with the periodic boundary condition to simulate a bulk Pt system undergoing liquid-solid phase transitions. This system does not have a surface constraint for nucleation, but its early stage nuclei remain nonspherical (Extended Data Fig. 12b), which is consistent with our experimental observations (Extended Data Fig. 4h).

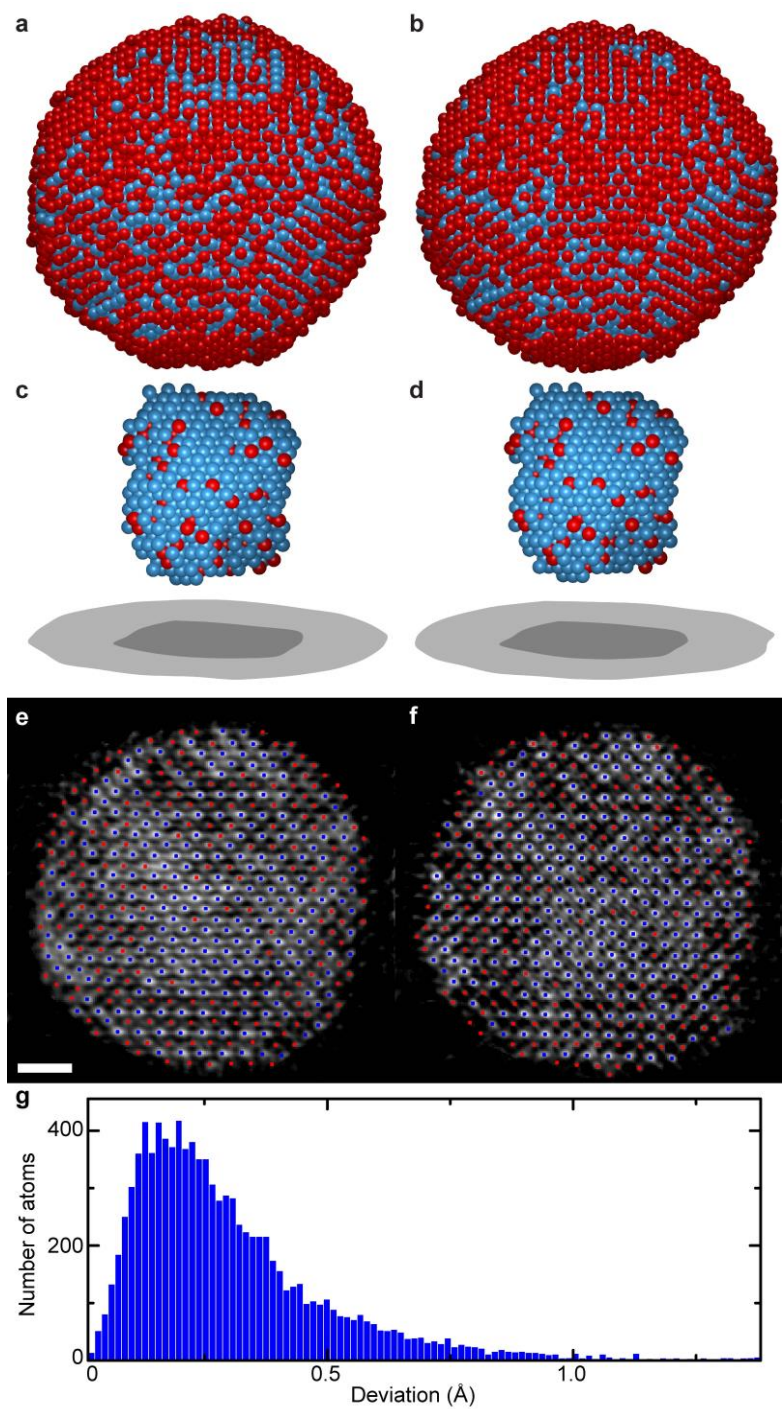

**Extended Data Figure 1 | Consistency check of the AET measurements. a** and **b**, 3D atomic models (Pt in blue and Fe in red) of particle 1, obtained from two independent experimental measurements. **c** and **d,** Pt-rich cores cropped from the atomic models shown in (**a**) and (**b**), respectively. **e** and **f,** The same atomic layer of the nanoparticle along the [010] direction (Pt in blue and Fe in red), obtained from the two independent measurements. Scale bar, 1nm. **g**, Histogram of the deviation of the common atoms between the two independent measurements. By dividing twice of the common atoms by the total number of atoms in the

two measurements, we estimated that 95.4% of the atoms are consistent. The average deviation between the two independent measurements is 37 pm. According to the statistical analysis of error propagation, the precision of the AET measurement is 37 pm $/\sqrt{2}$ = 26 pm.

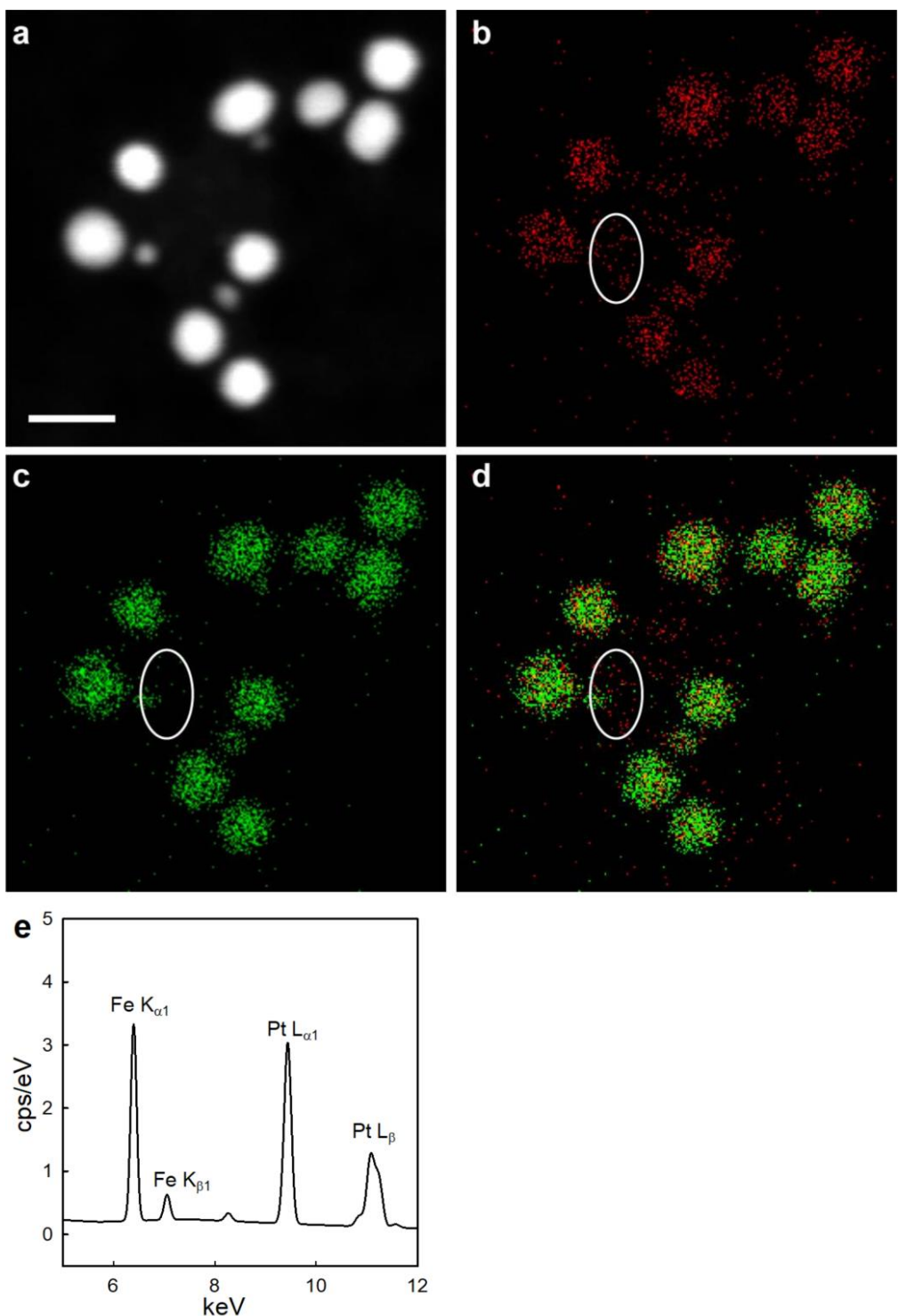

**Extended Data Figure 2 | Distribution of Fe and Pt atomic nanoclusters between FePt nanoparticles. a**, ADF-STEM image of the FePt nanoparticles on a $Si_3N_4$ substrate. Energy-dispersive x-ray spectroscopy (EDS) images show the distribution of Fe (**b**), Pt (**c**), and both Fe and Pt atomic nanoclusters (**d**) between FePt nanoparticles, acquired simultaneously with the ADF-STEM image (**a**). (**e**) Fitted spectrum of Fe (K-edges) and Pt (L-edges) from the ellipse region in (**d**), where cps stands for counts per second. Scale bar, 10 nm.

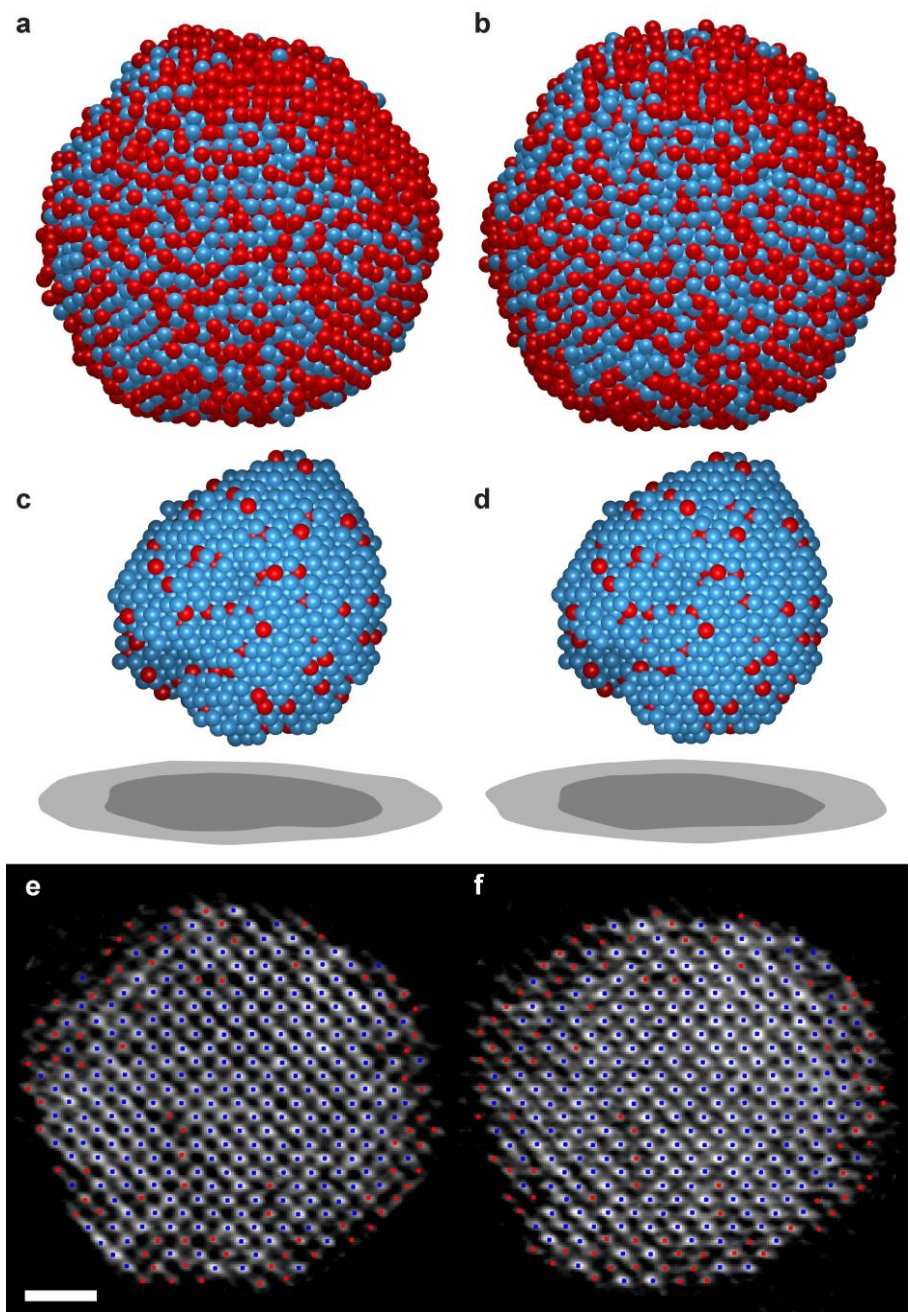

**Extended Data Figure 3 | 4D AET of an FePt nanoparticle at two annealing times. a** and
**b,** 3D atomic models (Pt in blue and Fe in red) of particle 3 with a total annealing time of 9
and 16 minutes, respectively, determined by AET. **c** and **d,** The Pt-rich core of the
nanoparticle remained the same between the two annealing times. The light and dark grey
projections show the whole nanoparticle and the core, respectively. **e** and **f,** The same atomic
layer of the nanoparticle along the [010] direction at the two annealing times (Pt in and Fe in
red), where a fraction of the surface and sub-surface atoms were re-arranged due to the
annealing process, but the Pt-rich core of the nanoparticle did not change. Scale bar, 1nm.

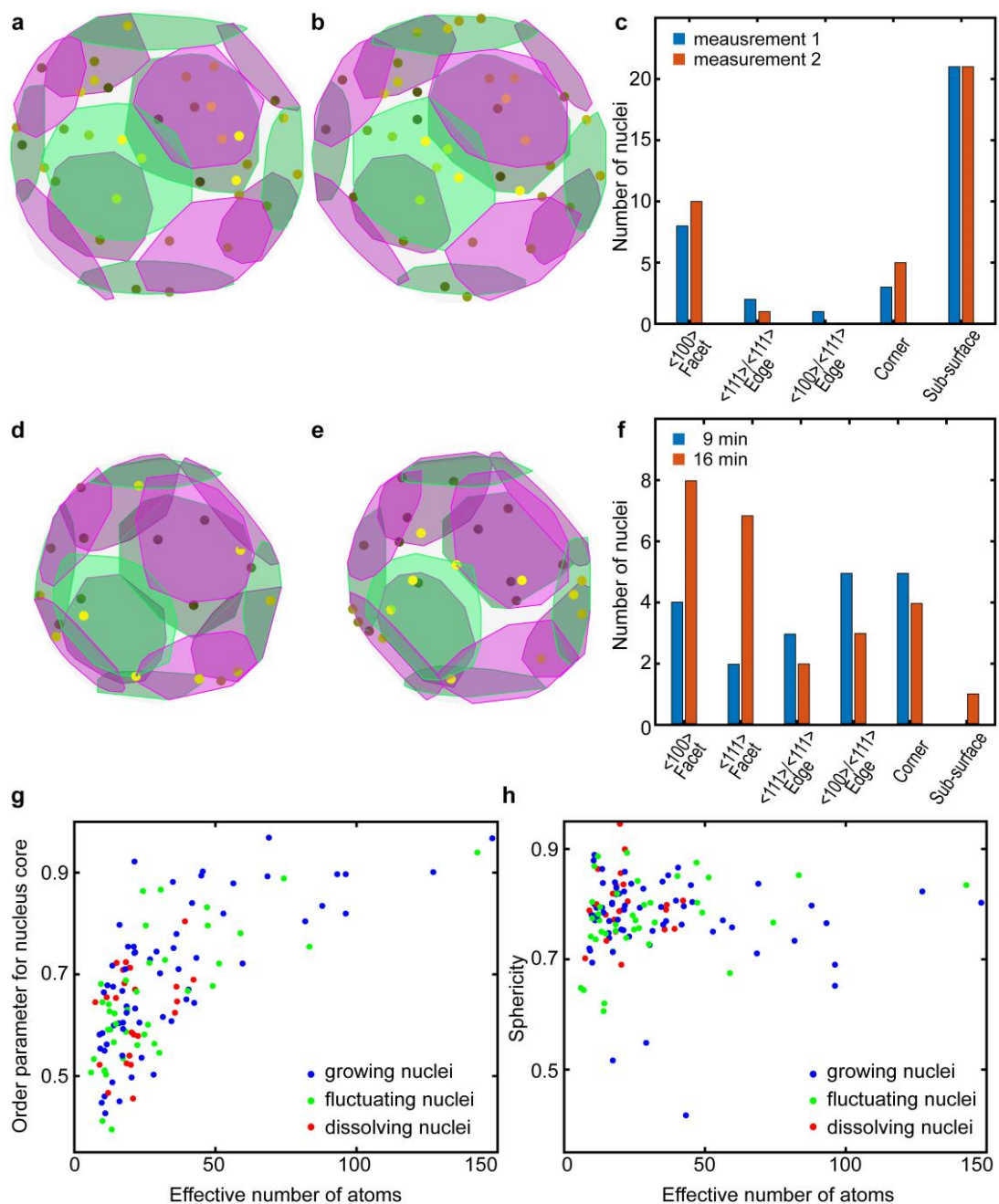

**Extended Data Figure 4 | Analysis of the sites, cores and 3D shapes of early stage nuclei.**
**a** and **b**, The distribution of the nucleation sites (circular dots) in particle 1 obtained from two independent measurements, where the lighter colour dots are closer to the front side and the darker dots are closer to the back side of the nanoparticle. The <100> and <111> facets are in magenta and green, respectively. **c**, Histogram of the nucleation site distribution in particle 1. Compared to particles 2 and 3, particle 1 has more nucleation sites at the sub-surface, because many nuclei in particle 1 are relatively large and their cores are more than one unit cell distance from the surface. **d** and **e**, The distribution of the nucleation sites (circular dots) in particle 3 with an annealing time of 9 and 16 minutes, respectively. **f**, Histogram of the nucleation site distribution in particle 3. **g**, The order parameter of the nucleus core as a function of the effective number of atoms for particles 2 and 3. **h**, The sphericity of the nuclei as a function of the effective number of atoms for particles 2 and 3.

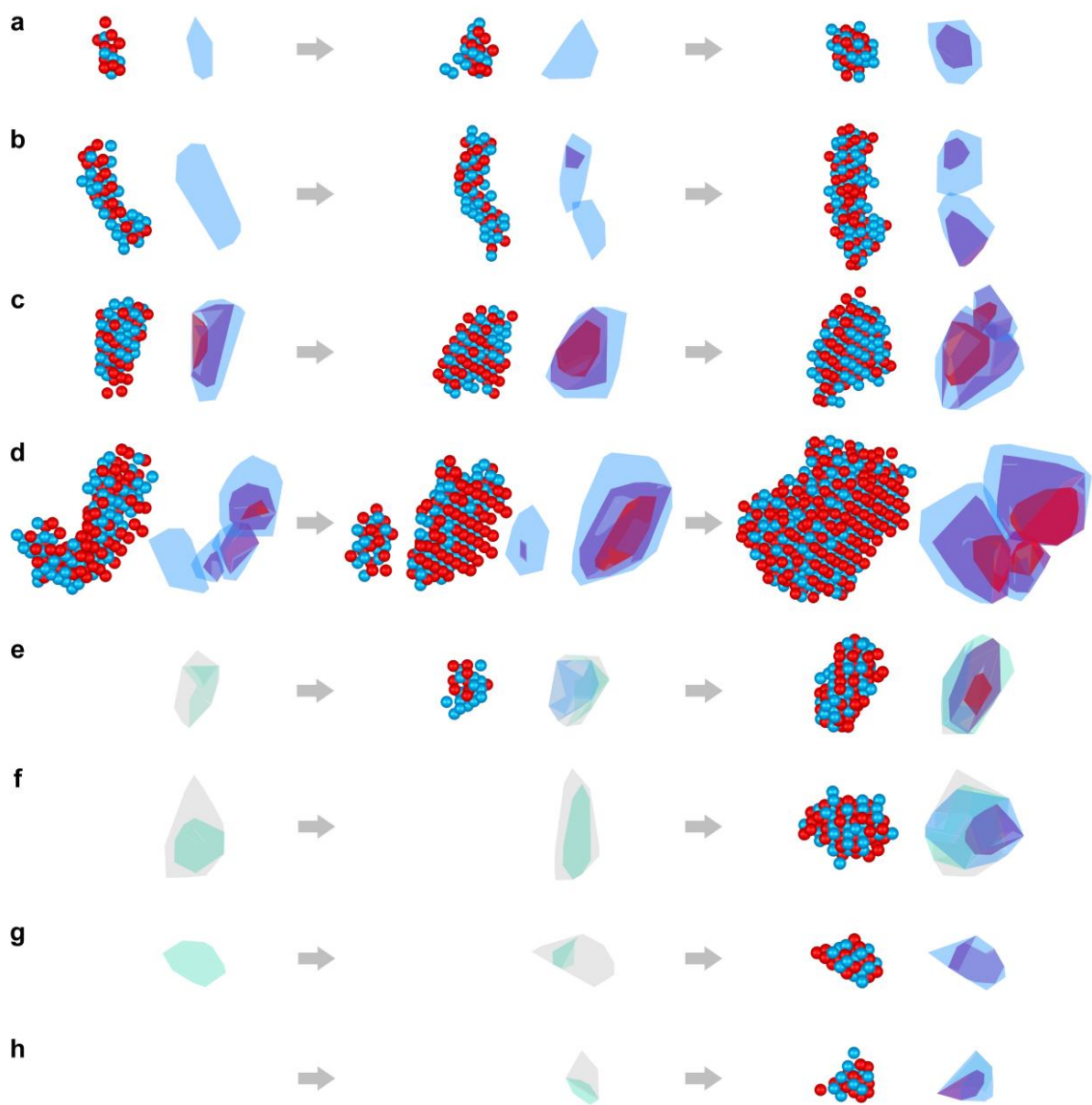

**Extended Data Figure 5 | Experimental observation of growing nuclei at 4D atomic resolution. a-d,** Four representative growing nuclei in particle 2 with a total annealing time of 9, 16 and 26 minutes, respectively, where the atomic models show Fe (red) and Pt atoms (blue) with an order parameter ≥ 0.3 and the 3D contour maps show the distribution of an order parameter of 0.7 (red), 0.5 (purple) and 0.3 (light blue). Dividing and merging nuclei are observed in (**b-d**). **e-h,** Another four representative growing nuclei in particle 2 with a total annealing time of 9, 16 and 26 minutes, where the 3D contour maps show the distribution of an order parameter of 0.7 (red), 0.5 (purple), 0.3 (light blue), 0.2 (green), and 0.1 (gray). No atomic model is displayed if a corresponding common nucleus was not identified at a specific annealing time. Another five growing nuclei in particle 3 similar to (**e-h**) are not shown here.

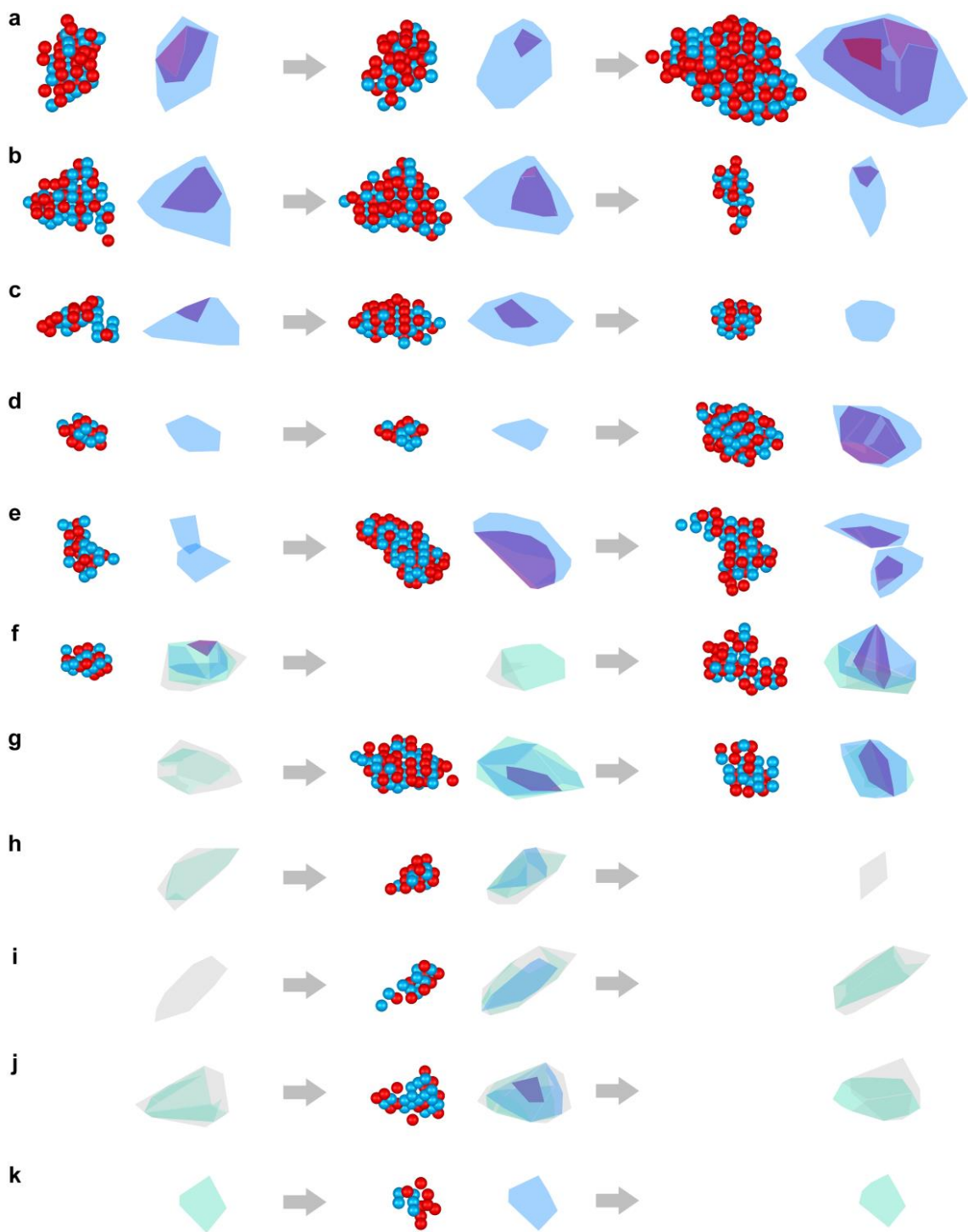

**Extended Data Figure 6 | Experimental observation of fluctuating nuclei at 4D atomic resolution. a-e**, Five representative fluctuating nuclei in particle 2 with a total annealing time of 9, 16 and 26 minutes, respectively, where the atomic models show Fe (red) and Pt atoms (blue) with an order parameter ≥ 0.3 and the 3D contour maps show the distribution of an order parameter of 0.7 (red), 0.5 (purple) and 0.3 (blue). Merging and dividing nuclei are observed in (**e**). **f-k**, Another six representative fluctuating nuclei in particle 2 with a total annealing time of 9, 16 and 26 minutes, where the 3D contour maps show the distribution of an order parameter of 0.7 (red), 0.5 (purple), 0.3 (light blue), 0.2 (green), and 0.1 (gray). No atomic model is displayed if a corresponding common nucleus was not identified at a specific annealing time.

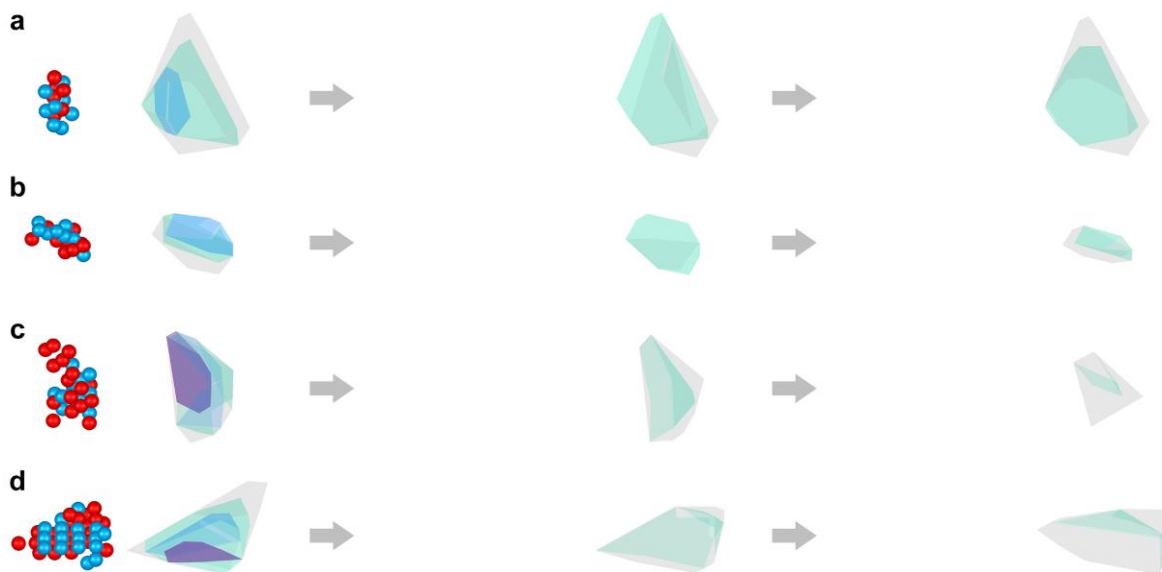

**Extended Data Figure 7 | Experimental observation of dissolving nuclei and schematic illustrations for the OPG nucleation model. a-d**, Four dissolving nuclei in particle 2 with a total annealing time of 9, 16 and 26 minutes, where the 3D contour maps show the distribution of an order parameter of 0.7 (red), 0.5 (purple), 0.3 (light blue), 0.2 (green), and 0.1 (gray). No atomic model is displayed if a corresponding common nucleus was not identified at a specific annealing time.

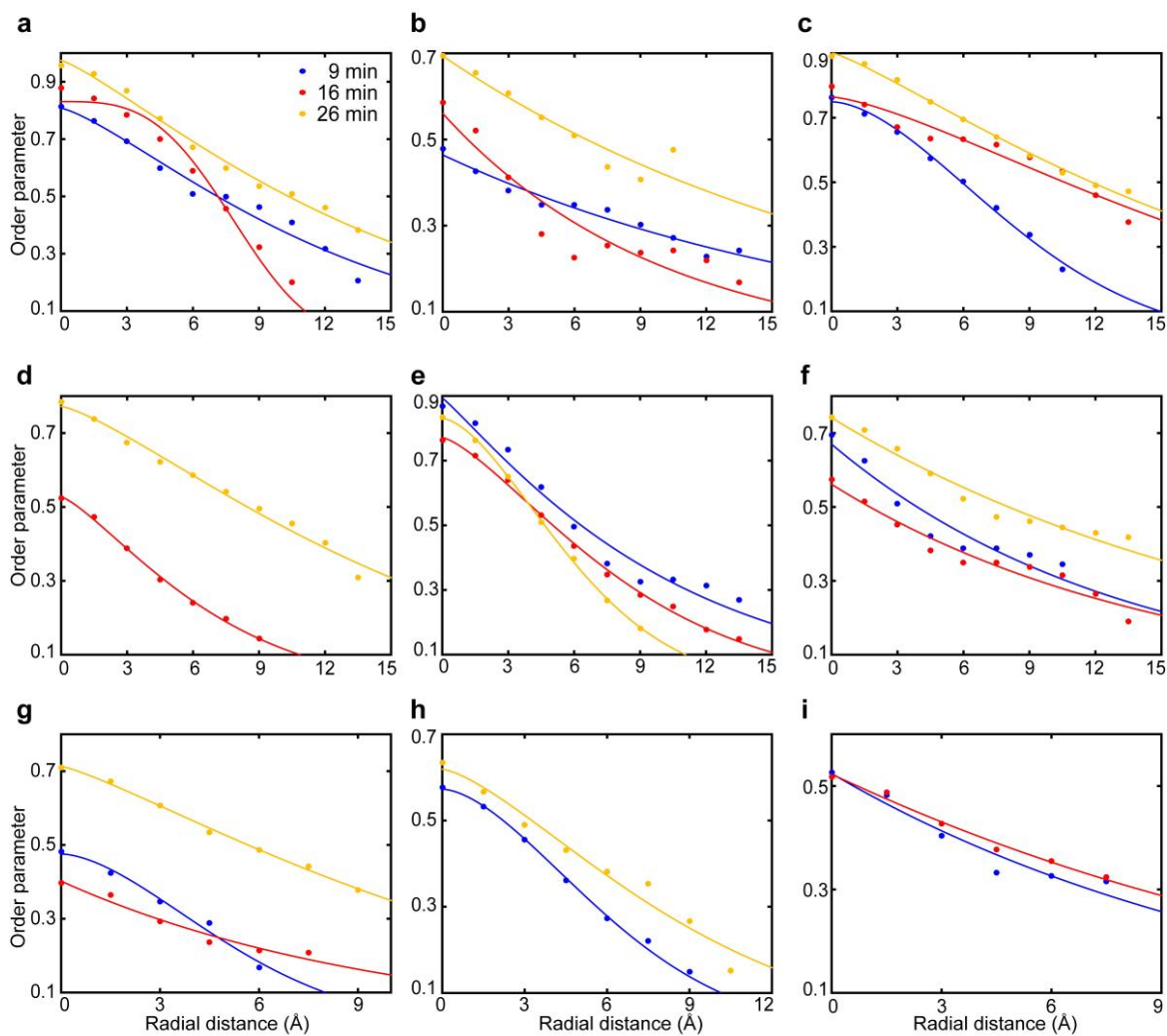

**Extended Data Figure 8 | Radial average order parameter distributions of nine representative nuclei.** The order parameter distributions for four growing nuclei (**a-d**), four fluctuating nuclei (**e-h**) and one dissolving nucleus (**i**) in particle 2, where the dots represent the experimentally measured data and the curves are fitted with equation (4).

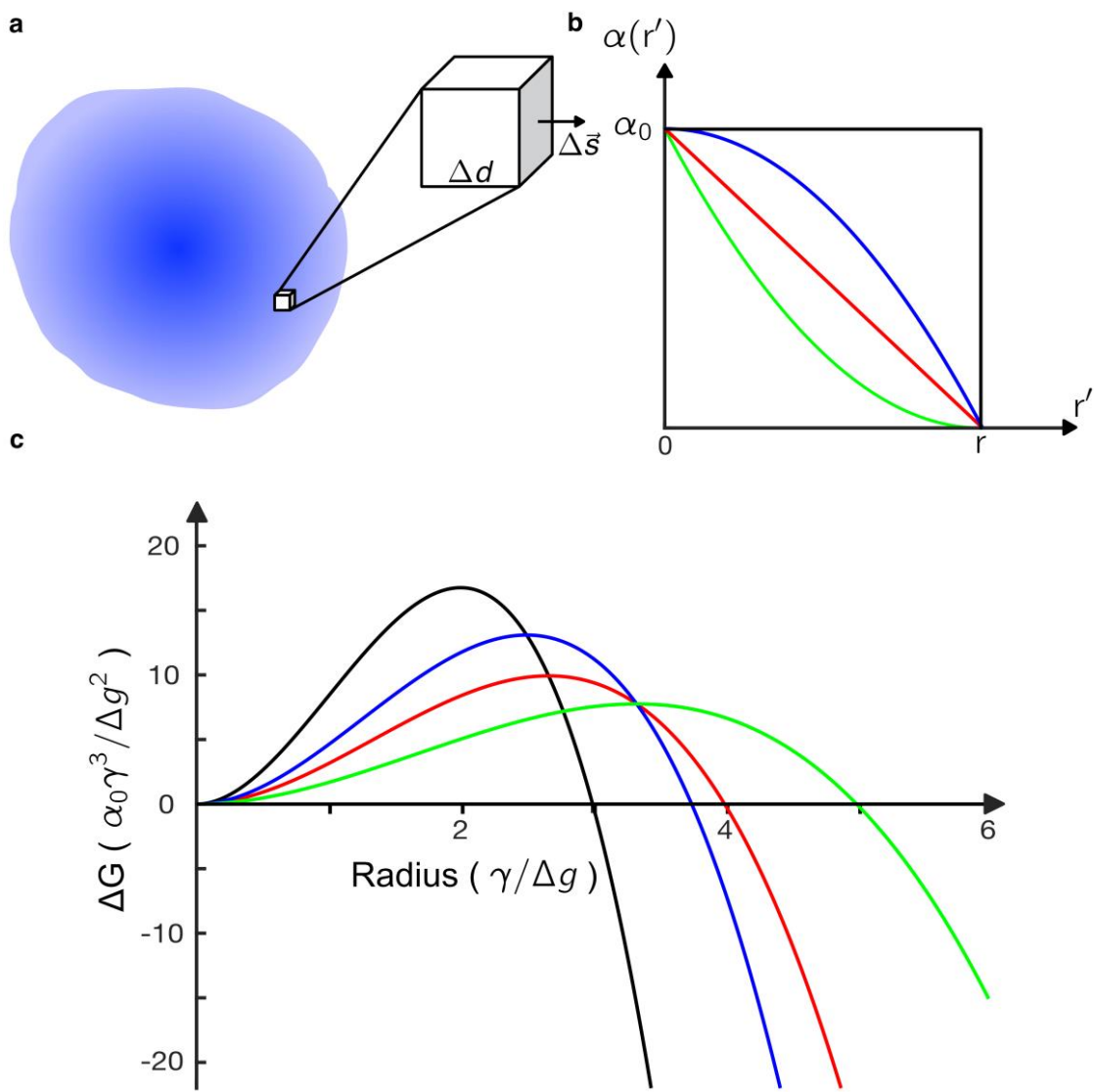

**Extended Data Figure 9 | Schematic illustration of the OPG model and its energy barriers. a**, Schematic illustration of a nucleus with varying order parameters. The nucleus is divided into many very small volumes for the calculation of the interfacial energy. **b**, Four specific cases of the order parameter distribution ($\alpha$): the Heaviside step function (black), a linear function (red), and two parabolic functions (blue and green). **c**, The total free energy change as a function of the radial distance calculated from the Heaviside step function (black), the linear function (red), and the two parabolic functions (blue and green) using the OPG model. With a small radius, the interfacial energy term (the second term in equation (1)) dominates, creating a lower energy barrier of the diffusive interface (the red, blue and green curves) than that of the sharp interface (the black curve). With a large radius, the volume energy change term (the first term in equation (1)) dominates, making the total free energy of the sharp interface (the black curve) decrease faster than those of the diffuse interface (the red, blue and green curves).

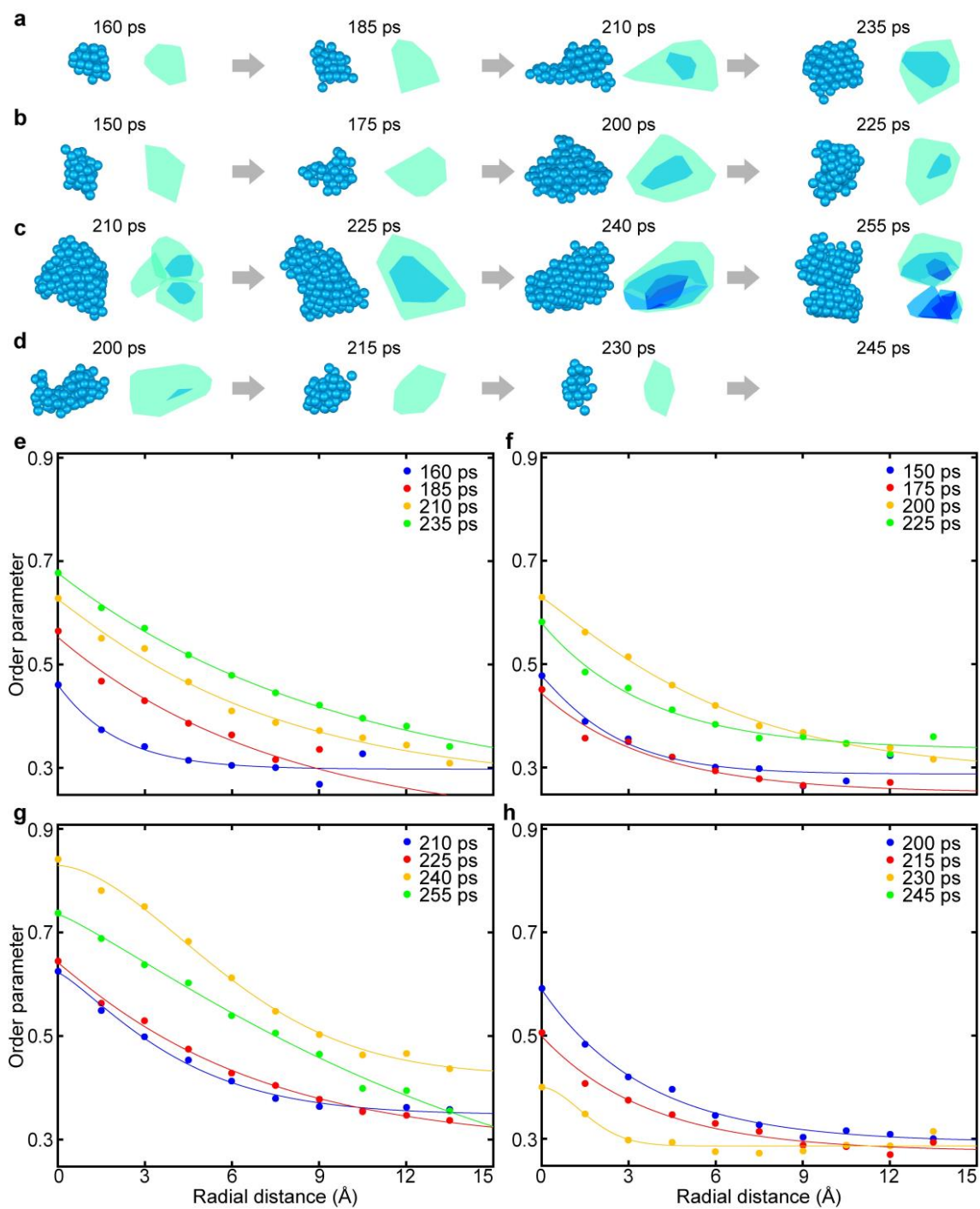

**Extended Data Figure 10 | Nucleation dynamics in the liquid-solid transition of a Pt nanoparticle, obtained by MD simulations with the interface force field**. **a**, A representative growing nucleus, where the atomic models show the Pt atoms with an order parameter ≥ 0.3 and the 3D contour maps show the distribution of an order parameter of 0.7 (dark blue), 0.5 (light blue) and 0.3 (cyan). **b** and **c**, Two representative fluctuating nuclei, where merging and dividing nuclei are observed in (**c**). **d**, A representative dissolving nucleus, which dissolved at 245 ps. **e-h**, Radial average order parameter distributions of the four nuclei shown in (**a-d**), respectively, where the dots were obtained by time-averaging ten consecutive MD snapshots with 1 ps time intervals and the curves are the fitted results using equation (4) with a constant background.

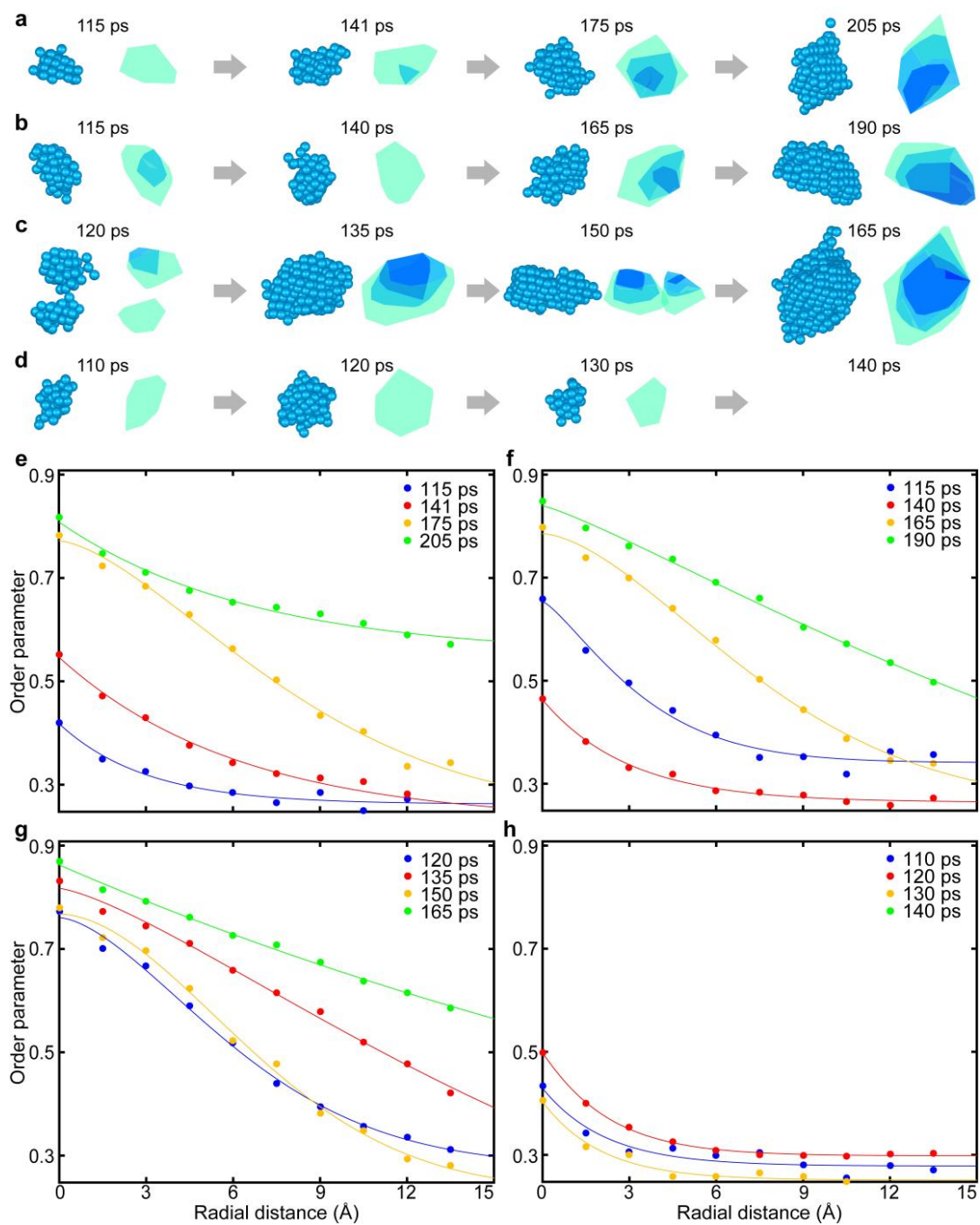

**Extended Data Figure 11 | Nucleation dynamics in the liquid-solid transition of a bulk Pt system, obtained by MD simulations with the embedded-atom method potential**. **a**, A representative growing nucleus, where the atomic models show the Pt atoms with an order parameter ≥ 0.3 and the 3D contour maps show the distribution of an order parameter of 0.7 (dark blue), 0.5 (light blue) and 0.3 (cyan). **b** and **c**, Two representative fluctuating nuclei, where merging and dividing nuclei are observed in (**c**). **d**, A representative dissolving nucleus, which dissolved at 140 ps. **e-h**, Radial average order parameter distributions of the four nuclei shown in (**a-d**), respectively, where the dots were obtained by time-averaging ten consecutive MD snapshots with 1 ps time intervals and the curves are the fitted results using equation (4) with a constant background.

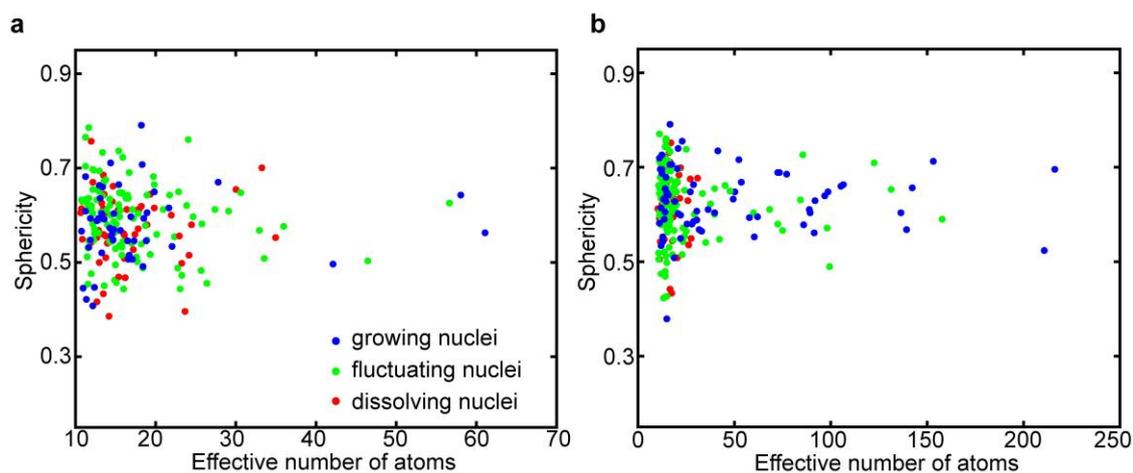

**Extended Data Figure 12 | Analysis of the 3D shapes of nuclei in liquid-solid phase transitions of Pt, obtained by MD simulations with the embedded-atom method potential**. **a** and **b**, The sphericity of the nuclei in a Pt nanoparticle (heterogeneous nucleation) and a bulk Pt system (homogeneous nucleation) as a function of the effective number of atoms, respectively.

**Extended Data Table 1 | AET data collection, reconstruction, refinement and validation statistics**

| | Particle 1 | | Particle 2 | | | Particle 3 | |
|---|---|---|---|---|---|---|---|
| | Tilt series #1 | Tilt series #2 | Tilt series #3 | Tilt series #4 | Tilt series #5 | Tilt series #6 | Tilt series #7 |
| **Data collection and processing** | | | | | | | |
| Annealing time (min) | 9 | 9 | 9 | 16 | 26 | 9 | 16 |
| Voltage (kV) | 200 | 200 | 200 | 200 | 200 | 200 | 200 |
| Convergence semi-angle(mrad) | 30 | 30 | 30 | 30 | 30 | 30 | 30 |
| Probe size (Å) | 0.7 | 0.7 | 0.7 | 0.7 | 0.7 | 0.7 | 0.7 |
| Detector inner angle (mrad) | 38 | 38 | 38 | 38 | 38 | 38 | 38 |
| Detector outer angle (mrad) | 190 | 190 | 190 | 190 | 190 | 190 | 190 |
| Depth of focus (nm) | 6 | 6 | 6 | 6 | 6 | 6 | 6 |
| Pixel size (Å) | 0.34 | 0.34 | 0.34 | 0.34 | 0.34 | 0.34 | 0.34 |
| # of projections | 57 | 55 | 52 | 52 | 51 | 52 | 52 |
| Tilt range (°) | −64.3 +65.3 | −64.3 +65.5 | −62.3 +63.1 | −62.3 +63.0 | −62.0 +62.1 | −62.3 +63.1 | −62.3 +63.0 |
| Electron dose ($10^5$ e/Å²) | 8.5 | 8.2 | 7.7 | 7.7 | 7.6 | 7.7 | 7.7 |
| **Reconstruction** | | | | | | | |
| Algorithm | GENFIRE | GENFIRE | GENFIRE | GENFIRE | GENFIRE | GENFIRE | GENFIRE |
| Interpolation method[a] | DFT | DFT | DFT | DFT | DFT | DFT | DFT |
| Interpolation radius (voxel) | 0.1 | 0.1 | 0.1 | 0.1 | 0.1 | 0.1 | 0.1 |
| Oversampling ratio | 4 | 4 | 4 | 4 | 4 | 4 | 4 |
| **Refinement** | | | | | | | |
| $R_1$ (%)[b] | 7.8 | 7.8 | 12.9 | 10.5 | 8.8 | 8.7 | 9.2 |
| $R$ (%)[c] | 20.8 | 20.0 | 14.7 | 17.0 | 17.9 | 15.8 | 15.4 |
| B' factors (Å²) | | | | | | | |
| Fe atoms | 24.3 | 22.3 | 25.4 | 25.0 | 23.0 | 23.4 | 23.5 |
| Pt atoms | 35.2 | 25.5 | 26.8 | 34.7 | 27.6 | 25.6 | 28.1 |
| # of atoms | | | | | | | |
| Fe | 5356 | 5407 | 1640 | 1773 | 2291 | 2103 | 2313 |
| Pt | 5107 | 5066 | 3195 | 3295 | 3195 | 4078 | 4127 |
| # of Common atoms | | | | | | | |
| Fe | 4996 | 4996 | 1375 | 1375/1383[d] | 1383 | 1805 | 1805 |
| Pt | 4986 | 4986 | 3090 | 3090/2808[e] | 2808 | 3880 | 3880 |

[a]GENFIRE uses either the discrete Fourier transform or the fast Fourier transform to obtain the Fourier coefficients. The former is slower but more accurate than the latter. [b]The $R_1$-factor is defined as equation (5) in ref. 36. [c]The R-factor is defined as $R = \frac{\sum ||F_{obs}| - |F_{cal}||}{\sum |F_{obs}|}$, where $|F_{obs}|$ is the Fourier magnitude obtained from experimental data and $|F_{cal}|$ the Fourier magnitude calculated from an atomic model. [d]1375 and 1383 are the common Fe atoms between tilt series #3 and #4 and between tilt series #4 and #5, respectively. [e]3090 and 2808 are the common Pt atoms between tilt series #3 and #4 and between tilt series #4 and #5, respectively.